\documentclass[journal,comsoc]{IEEEtran}

\usepackage{cite}
\usepackage[pdftex]{graphicx}
\usepackage{url}
\usepackage{siunitx} 
\usepackage{amsmath,amsfonts,amssymb,amsthm}
\usepackage{mathrsfs}

\usepackage[tight,footnotesize]{subfigure}

\usepackage[table]{xcolor}
\definecolor{headerColor}{rgb}{0.75, 0.9, 0.9}

\DeclareMathOperator*{\margmax}{\text{argmax}}

\theoremstyle{plain}

\theoremstyle{definition}
\newtheorem{defn}{Definition}

\definecolor{morange}{rgb}{0.8,0.2,0}
\definecolor{mblue}{rgb}{0,0,0}
\definecolor{mgreen}{rgb}{0.2,0.4,0}
\definecolor{mpurple}{rgb}{0.5 0.1 0.7}

\newcommand{\tabCrefA}{\cite{Kuran2010_Energy, Kuran2011_Modulation, Mahfuz2011_Characteristics} }

\newcommand{\tabCrefC}{\cite{Llatser2013_Detection} }
\newcommand{\tabCrefD}{\cite{Mahfuz2013_GeneralizedSB, Mahfuz2015_Comprehensive, Mahfuz2016_Concentration} }
\newcommand{\tabCrefE}{\cite{Singhal2014_Molecular, Singhal2015_Performance} }
\newcommand{\tabCrefF}{\cite{Einolghozati2011_Capacity, Tepekule2014_Energy, Movahednasab2015_Adaptive} }
\newcommand{\tabCrefFA}{\cite{Turan2018_MOL} }
\newcommand{\tabCrefG}{\cite{Kilinc2013_Receiver, Li2016_LComplexity} }
\newcommand{\tabCrefGA}{\cite{Damrath2016_LowComp} }
\newcommand{\tabCrefH}{\cite{Li2016_LComplexity} }
\newcommand{\tabCrefI}{\cite{Li2016_LocalC} }
\newcommand{\tabCrefJ}{\cite{Jamali2016_NonCoherent} }  
\newcommand{\tabCrefK}{\cite{Akdeniz2020_Equilibrium} }  

\newcommand{\tabTrefA}{\cite{Kuran2011_Modulation,Kuran2012_Interference,Aminian2015_Capacity, Galmes2016_Performance}}
\newcommand{\tabTrefB}{\cite{Kim2012_Novel, Kim2014_SymbolIO}}
\newcommand{\tabTrefC}{\cite{ShahM2012_Optimum}}
\newcommand{\tabTrefD}{\cite{Kim2013_Novel}}
\newcommand{\tabTrefE}{\cite{Arjmandi2013_Diffusion}}
\newcommand{\tabTrefF}{\cite{Tepekule2014_Energy}}
\newcommand{\tabTrefG}{\cite{Tepekule2015_Novel}}
\newcommand{\tabTrefH}{\cite{Pudasaini2014_Robust}}
\newcommand{\tabTimerefA}{\cite{Srinivas2012_Molecular,Li2014_Capacity}}
\newcommand{\tabTimerefB}{\cite{Eckford2007_Nanoscale,Eckford2008_Molecular,Rose2016_InscribedPartI,Rose2016_InscribedPartII}}
\newcommand{\tabTimerefC}{\cite{Farsad2016_CapacityOD,Farsad2019_CapacityLimits}}
\newcommand{\tabTimerefD}{\cite{Garralda2011_Diffusion,Akdeniz2018_PPM}}
\newcommand{\tabTimerefE}{\cite{zare2019receiverDA,Murin2019_OneShotPPM, Murin2018_PPMOrderStat}}
\newcommand{\tabTimerefF}{\cite{Murin2017_SortDetect}}
\newcommand{\tabTimerefG}{\cite{Garralda2011_Diffusion,Hsieh2013_Asynch, Krishnaswamy2013_Time-Elapse, Farsad2017_Communication}}
\newcommand{\tabTimerefH}{\cite{Mahfuz2010_Spatiotemporal, Chou2012_MolecularCF, Yilmaz2014_SimulationSO, Guo2015_MolecularLCW}}

\newcommand{\xth}[1]{{#1}\mbox{-th}}
\newcommand{\etal}{\text{et al.}}

\DeclareSIUnit{\pH}{pH}

\begin{document}

\title{A Survey on Modulation Techniques in Molecular Communication via Diffusion}

\author{Mehmet~\c{S}\"{u}kr\"{u}~Kuran,~\IEEEmembership{Member,~IEEE,}
        H.~Birkan~Yilmaz,~\IEEEmembership{Member,~IEEE,}
        Ilker~Demirkol,~\IEEEmembership{Senior Member,~IEEE,}\\
        Nariman~Farsad,~\IEEEmembership{Member,~IEEE,}
        and~Andrea~Goldsmith,~\IEEEmembership{Fellow,~IEEE}
\thanks{M. S. Kuran is with the Department
of Computer Engineering, Bahcesehir University, Istanbul,
Turkey, (e-mail: mehmetsukru.kuran@eng.bau.edu.tr)}
\thanks{H. B. Yilmaz is with the Computer Networks Research Laboratory (NETLAB), Dept. of Computer Engineering, Bogazici University, Istanbul, Turkey. (e-mail: birkan.yilmaz@boun.edu.tr)}
\thanks{I. Demirkol is with the Dept. of Mining, Industrial and ICT Engineering, Universitat Polit\`ecnica de Catalunya, Barcelona, Spain. (e-mail: ilker.demirkol@upc.edu)}
\thanks{Nariman Farsad is with the Dept. of Computer Science, Ryerson University, Toronto, ON M5B 2K3, Canada. (e-mail: nfarsad@ryerson.edu)}
\thanks{Andrea Goldsmith is with the Dept. of Electrical Engineering, Princeton University, Princeton, NJ 08540 USA. (e-mail: goldsmith@princeton.edu)}
\thanks{This research is supported in part by the Scientific and Technical Research Council of Turkey (TUBITAK) under BIDEB-2232 program with the grant number 118C274, NSERC Discovery under Grant RGPIN-2020-04926, the NSF Center for Science of Information (CSoI) under grant CCF-0939370, and CFI John Evans Leaders Funds. }
}

%
%

\markboth{Kuran \MakeLowercase{et al.}: Survey on Modulation Techniques in MCvD}%
{Kuran \MakeLowercase{et al.}: Survey on Modulation Techniques in MCvD}
%

\maketitle

\begin{abstract}
This survey paper focuses on modulation aspects of molecular communication, an emerging field focused on building biologically-inspired systems that embed data within chemical signals. The primary challenges in designing these systems are how to encode and modulate information onto chemical signals, and how to design a receiver that can detect and decode the information from the corrupted chemical signal observed at the destination. In this paper, we focus on modulation design for molecular communication via diffusion systems. In these systems, chemical signals are transported using diffusion, possibly assisted by flow, from the transmitter to the receiver. This tutorial presents recent advancements in modulation and demodulation schemes for molecular communication via diffusion. We compare five different modulation types: concentration-based, type-based, timing-based, spatial, and higher-order modulation techniques. The end-to-end system designs for each modulation scheme are presented. In addition, the key metrics used in the literature to evaluate the performance of these techniques are also presented. Finally, we provide a numerical bit error rate comparison of prominent modulation techniques using analytical models. We close the tutorial with a discussion of key open issues and future research directions for design of molecular communication via diffusion systems.
\end{abstract}


%
\IEEEpeerreviewmaketitle

\section{Introduction}
\IEEEPARstart{M}{olecular} communication (MC) is an emerging area that relies on chemical signals for transferring information from a transmitter to a receiver \cite{Akyildiz2008_Nano, nakano2012_Molecular, nakano_eckford_haraguchi_2013, Farsad2016_Comprehensive}. Since MC systems have different characteristics compared to traditional communication systems that embed data into electromagnetic (EM) signals, MC systems can be used in areas where EM communication fails or is not feasible. For example, they can be used for communication inside complex networks of metallic ducks and pipes \cite{Gine2009_LongRange,Qui2014_ConfinedDuct}, where wireless signals fail. Another benefit of MC compared to EM signalling is energy efficiency. In particular, although MC cannot achieve the data rates that are obtained by EM-based systems, they have much lower energy spent per transmitted information bit \cite{Kuran2010_Energy,Rose2019_CapacityIEEEProc}, making them suitable for ultra low power applications that do not require high data rates. 

MC can also be used to connect tiny {\em engineered} devices. Specifically, recent advances in the fields of bio-engineering and nanotechnology have resulted in the emergence of tiny devices of sub-millimeter dimensions that can perform sensing and actuation. For example, synthetic cells are excellent bio-marker sensors and can detect bio-markers for cancer cells {\em in vivo} at small concentrations \cite{anderson2006_CancerDetect,danino2015_ProgrammableBacteria, slomovic2015_SyntheticBacteria}. As another example, micro-sized devices based on graphene could be used for removal of nano-sized toxic contaminants \cite{zhu2015_PrintedFish, vilela2016_GrapheneDevice}. Although these devices have been shown to work in a laboratory setting, moving them out of the laboratory and into practical settings remains an open challenge. In particular, for many applications, these tiny devices need to communicate and collaborate in swarms, or they need to transmit their measurements to other devices (such as micro-sized sink nodes or health monitoring wristwatches). Since chemical signaling is already used in many biological systems to interconnect cells or regulate bodily functions, MC is a promising solution for interconnecting tiny devices, due to its energy efficiency and bio-compatibility (i.e., synthetic biological devices are well-suited to transmit and receive chemical signals with appropriate enhancement of their communication capability.)
\begin{figure*}[t]
  \centering
    \includegraphics[width=1.99\columnwidth]{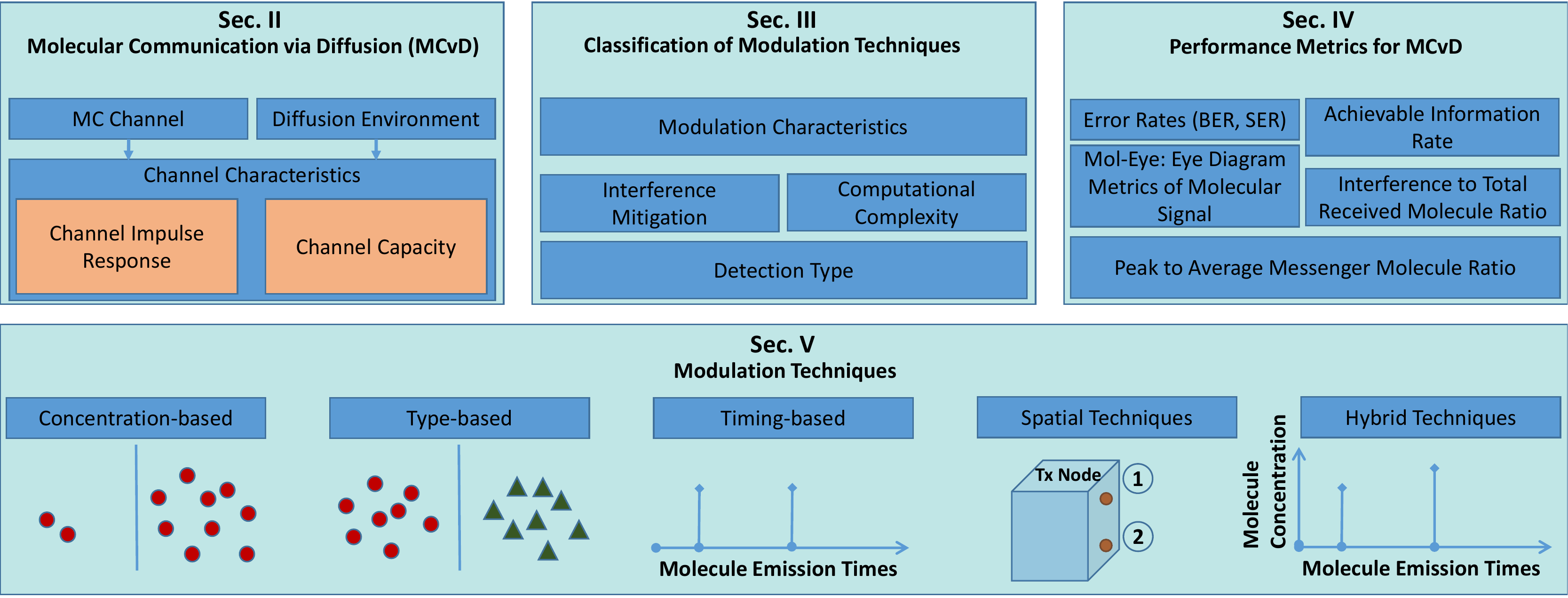}
  \caption{Organization and contents of the paper and its sections.}
  \label{fig:paper_structure}
\end{figure*}

Different types of MC systems have been investigated in the literature such as molecular communication via diffusion (MCvD) \cite{Suda2005_Exploratory, nakano_eckford_haraguchi_2013, Kuran2010_Energy}, bacteria-based communication \cite{Gregori2010_Nanonetwork}, microtubule-based communication \cite{Moore2006_Design, Darchini2013_Molecular, Chahibi2016_Propagation}, calcium signaling \cite{Nakano2007_Molecular, Nakano2008_Molecular, Kuran2012_Calcium}, and pheromone signaling \cite{Gine2009_LongRange, Unluturk2017_ETE}. 
Among these MC systems, the MCvD has emerged as the technology of choice due to its versatility and applicability to many environments. The communication media and signalling scheme utilized by MCvD system are vastly different from that of classical communication systems. The first step towards design of these systems is modeling the MCvD system components including channel, transmitter, receiver, signal, noise, and interference. In recent years, several key survey and tutorials on such works have been published as follows:
\begin{itemize}
\item Farsad \etal{} present an overall look at the recent advances in the greater topic of molecular communication \cite{Farsad2016_Comprehensive}.
\item Jamali \etal{} investigate and review various works and approaches on modeling the molecular channel \cite{Jamali2018_Channel}.
\item Kuscu \etal{} elaborate and present different approaches on transmitter and receiver architectures for molecular communication \cite{Kuscu2019_Transmitter}.
\end{itemize}

In addition to the modeling of these components, a wide range of modulation techniques have also been proposed in the MCvD literature utilizing different aspects of the molecular signal and the aforementioned components of the MCvD system. These techniques aim to increase the overall performance of the communication and are another key part of the overall MCvD design. Although the previously mentioned tutorials briefly include some well-known modulation techniques, they do not provide a detailed and categorized look at the plethora of modulation techniques that have emerged in recent years. In contrast, this paper focuses on and presents a survey of these modulation techniques.


Since prior works have been based on different system assumptions, the performances of the proposed techniques cannot be directly compared with one another. To overcome this challenge, in our approach, we categorize and group prior works according to the characteristics of the modulation scheme and the assumptions made for their performance evaluation. We consider three key characteristics: inter-symbol interference (ISI) mitigation, computational complexity, and modulation type. Then, using this categorization, we offer a survey of various MCvD modulation techniques. We also provide a quantitative comparison between the most prominent techniques by comparing their bit error rate performance using analytical models. Finally, we underline and briefly elaborate on the current open problems related to the modulation technique design of MC.

The main contributions of this paper are summarized as follows:
\begin{itemize}
\item We give a comprehensive discussion on the various performance evaluation metrics being used in the literature to evaluate MCvD systems.
\item We present a detailed survey of the various modulation techniques that have been proposed for the MC system.
\item We provide a systematic approach to compare different modulation techniques, including current as well as future methods.
\item We perform a case study on the most prevalent modulation techniques where we evaluate the bit error rate performances of each technique and provide a performance comparison of them under the same conditions.
\end{itemize}

The remainder of the paper is organized as follows, which is also depicted in Fig.~\ref{fig:paper_structure}. In Section \ref{sec:Classification}, a general overview of MC systems is presented. Section~\ref{sec:Classification} describes the classification approach for investigating MC modulation techniques. Section~\ref{sec:Perf_Metrics} describes a variety of performance metrics used in the literature to evaluate the performance of MC systems. In Section \ref{sec:Mod_Tech}, a detailed survey of the modulation techniques that have been proposed for MC systems is presented. Section \ref{sec:Open_Issues} elaborates on the key open problems and challenges on the design of modulation techniques for MC. Finally, Section \ref{sec:Conclusion} concludes the paper.
\begin{figure*}[t]
  \centering
    \includegraphics[width=1.5\columnwidth]{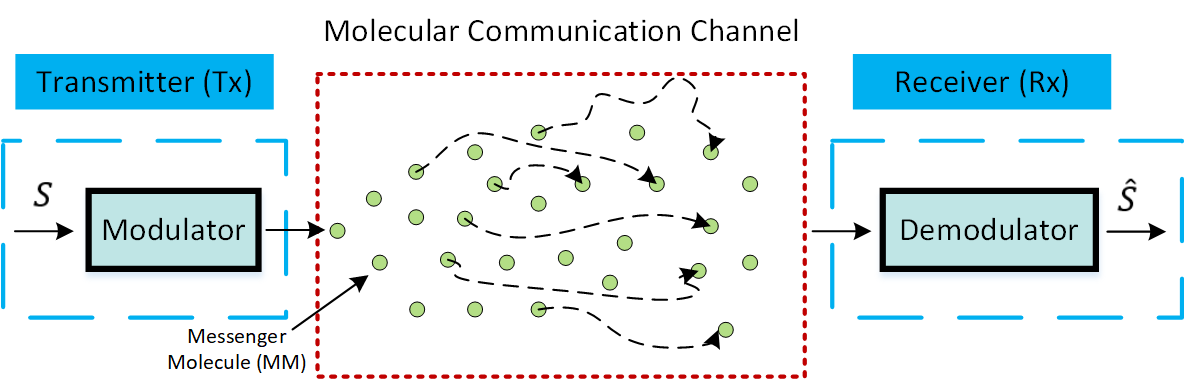}
  \caption{MCvD system and its three core components: the transmitter (Tx), the molecular communication channel, and the receiver (Rx).}
  \label{fig:mol_com_via_diffusion}
\end{figure*}

\section{\label{sec:Classification}Molecular Communication via Diffusion}

We consider an MCvD system as shown in Fig.~\ref{fig:mol_com_via_diffusion} whereby a sequence of input symbols (or bits) are modulated in a time slotted manner into $M$ symbols
\begin{equation} \label{eq:S_input}
\mathbf{S} = [S_1, S_2, ... , S_M],
\end{equation}
where $S_k \in \mathcal{S}$ refers to the $\xth{k}$ symbol that is modulated onto a chemical signal by encoding the symbol into some properties of the chemical emission process, and $\mathcal{S}$ is the symbol set. The modulated signal is transmitted through the MC channel, where the propagation environment degrades the signal and introduces delay in a probabilistic manner. The channel impaired chemical signal is detected and demodulated at the receiver. Let 
\begin{equation} \label{eq:S_output}
\hat{\mathbf{S}} =  [\hat{S}_1, \hat{S}_2, ... , \hat{S}_M], 
\end{equation}
denote the set of received symbols such that $\hat{S}_k$ refers to the \xth{k} symbol demodulated and detected at the receiver. The detected symbols are fed to the demodulator that can correct some of the errors in detection to recover the information.

At its core, the MCvD system utilizes small molecules or information particles called messenger molecules (MM) to relay information between the transmitter (Tx) and the receiver (Rx) over the MC channel. Due to the time slotted manner of the overall system, the Tx and Rx have to be synchronized with each other throughout the communication. Here we follow the general MCvD literature and assume that Tx and Rx are synchronized with each other using a synchronization method which is out-of-the-scope of this paper. MMs are generally a few nanometers to a few micrometers in size and can be composed of proteins, ions, magnetic nano particles, etc. The Tx chemically produces MMs, encodes and modulates the information over one or more physical properties of an MM signal, and finally releases these MMs to the channel. Unless stated otherwise, we assume that the MMs are released at the start of the symbol slot. Within the MC channel, due to their small size, the movement of MMs are chiefly governed by a type of diffusion process that depends on the particular features of the channel (e.g., free diffusion or diffusion with drift). 

Over time, some of these MMs reach the Rx, which uses a detection process that depends on the type of the MM and the type of the receiver used for detection. The detection processes that dominate the MC literature to date are passive receivers and absorbing receivers. In passive receivers, the Rx acts like a transparent entity (i.e., as if the Rx is not in the environment) and does not affect the movement of the MMs. The detection consists of counting the concentration of the MMs inside the Rx at certain time intervals. In absorbing receivers, the Rx acts as an absorbing entity for the MMs and when the MMs hit the Rx, they are removed from the environment. In these receivers, the detection can be modeled as counting the number of the MMs that hit the Rx within a given time interval. A sub-category of the absorbing receivers is the partially absorbing receivers, where only some parts of the receiver can absorb MMs while other parts simply reflect them \cite{akkaya2015effectOR}. 

\subsection{Molecular Communication Channel}
\label{subsec:mc_channel_properties}
A key aspect of the MCvD system is the diffusion process of the MC channel, which is vastly different from the electromagnetic wave propagation in wired and wireless communication channels. In the MCvD channel, movement of a single molecule is governed by Brownian motion, which is modelled by the Wiener process\footnote{Wiener process is also called Standard Brownian motion.}. When a group of particles collectively exhibit Brownian motion, the resulting process is called a diffusion process. A detailed mathematical foundation for the diffusion-based MC channel has been given by Hsieh \etal{} in \cite{Hsieh2013_Robust}, where it has been shown that the diffusion-based MC channel is a stationary and ergodic channel with memory. 

The diffusion process can be simulated via Monte Carlo simulations. In a 3-dimensional (3D) free diffusion MC channel, the displacements at each dimension, over a small time duration $\Delta{t}$, are independent and identically distributed (iid) random variables and can be numerically simulated as:
\begin{align}\label{eq:displacement_free}
\begin{split}
    \overrightarrow{r}[k] &= \overrightarrow{r}[k\!-\!1] + \Delta \overrightarrow{r}, \\
    \Delta \overrightarrow{r} &= (\Delta r_x,\, \Delta r_y,\, \Delta r_z), \\
    \Delta{r_i} &\sim N(0, 2D\Delta{t}) \quad \forall(i) \in \{x,y,z\},
\end{split}
\end{align}
where $\overrightarrow{r}[k]$ is the location of the MM at the \xth{k} time instance, $r_i$ represents $\xth{i}$ component of the location vector $\overrightarrow{r}$, $\Delta{t}$ is the discrete simulation time step, $D$ is the diffusion coefficient, and ${N\text{(}\mu\text{, }\sigma^{\text{2}}\text{)}}$ is the Gaussian random variable with mean $\mu$ and variance $\sigma^2$. 


In the MC channel, the movement speed of MMs are chiefly governed by the diffusion coefficient ($D$). This coefficient is affected by the temperature of the environment, $T$, the dynamic viscosity of the fluid which the MMs diffuse in, $\eta$, as well as the size of the MMs via their Stoke's radius, $r_{s}$ \cite{Harris1984_diffusion}. In particular, $D$ is given by the Stokes-Einstein relation as: 
\begin{equation}\label{eq:D}
D= \frac{k_{B}T}{\alpha\pi\eta r_{s}}
\end{equation}
where $k_{B}= 1.38 \cdot 10^{-23}$ J/K is the Boltzman constant and $\alpha$ is a unitless quantity whose value is mainly governed by the coefficient of sliding friction between the MMs and the molecules in the fluid, denoted as $\beta$. The SE relation is derived to describe the diffusive motion of molecules within a fluid composed of smaller particles \cite{Cappelezzo2007_Stokes}. Therefore, $\beta$ has two limits: “$\infty$” referring to a “stick” boundary and “$0$” referring to a “slip” boundary \cite{Costigliola2019_Revisiting}. The “stick” boundary refers to the case where size of the MMs ($s_{mm}$) are considerably bigger than the size of the molecules in the fluid ($s_{fluid}$) which yields $\alpha = 6$. On the other hand, the “slip” boundary refers to the case where $s_{mm}\approx s_{fluid}$ yielding $\alpha = 4$. Note that these are the two boundary conditions and depending on the relative size of $s_{mm}$ and $s_{fluid}$, $\alpha$ can also take other values. In this work, we follow the general diffusion literature and select $\alpha = 6$ since in a classical MC environment $s_{mm}$ is greater than $s_{fluid}$ such as an insulin molecule diffusing inside water (i.e., $s_{mm}=\SI{2.5}{nm}$ and $s_{fluid}=\SI{0.19}{nm}$).

As the value of $D$ depends on the properties of both the MMs and the environment (e.g., temperature of the fluid), assuming constant $D$ means that the change in the environment is negligible. Such scenarios are called, ideal diffusion scenarios. On the other hand, $D$ can be  modeled as a time-varying, space-varying, and/or molecular-density-varying value (e.g., the diffusion coefficient can change as the concentration of MMs in the environment increases). In such channels, the ideal diffusion is not applicable and more complex diffusion processes are required \cite{Jamali2018_Channel}. In this survey we only consider channels with constant $D$ values.

Another MC channel variant is the MCvD channel with flow. In such a channel, in addition to the diffusion process, a flow also affects the movement of the MMs. Considering an MCvD channel with flow only in the x dimension, \eqref{eq:displacement_free} becomes:
\begin{align}\label{eq:displacement_flowX}
\begin{split}
    \overrightarrow{r}[k] &= \overrightarrow{r}[k\!-\!1] + \Delta \overrightarrow{r}, \\
    \Delta \overrightarrow{r} &= (\Delta r_x^{d}+\Delta r_x^{flow},\, \Delta r_y^{d},\, \Delta r_z^{d}), \\
    \Delta{r_i^{d}} &\sim N(0, 2D\Delta{t}) \quad \forall(i) \in \{x,y,z\},
\end{split}
\end{align}
where $\Delta{r_i^{d}}$ represents the movement due to diffusion and $\Delta{r_i^{flow}}$ represents the movement due to flow.

The effect of the flow depends on the type of flow considered in the channel model as well as the diffusion environment. Considering a closed environment such as a pipe, flow could either be laminar, i.e. following constant streamlines, or turbulent depending on the velocity and viscosity of the fluid in the channel. Laminar flow occurs at lower velocities, whereas turbulent flow occurs at higher velocities. This fluid velocity is determined by a dimensionless parameter called the Reynolds number, which is the ratio of the inertial forces in a fluid and the viscous forces. Generally, studies that focus on MC channels with flow consider laminar flow only.

\subsection{Diffusion Environment}

The diffusion environment that is considered in the previous subsection only consists of the Tx, Rx, and the MMs used in the communication. More complex environments consider additional physical objects or barriers that limit the overall communication in the channel.

A constrained diffusion environment is a generalized version of the free diffusion environment that includes environmental objects other than the Tx and the Rx. These objects can be reflective, absorbing, or partially-absorbing to the MMs. Although more realistic, mathematical analysis of constrained diffusion environments is much more challenging than that of the free diffusion environments due to their inherent geometrical asymmetry.

\begin{figure}[t]
  \centering
    \includegraphics[width=1.0\columnwidth]{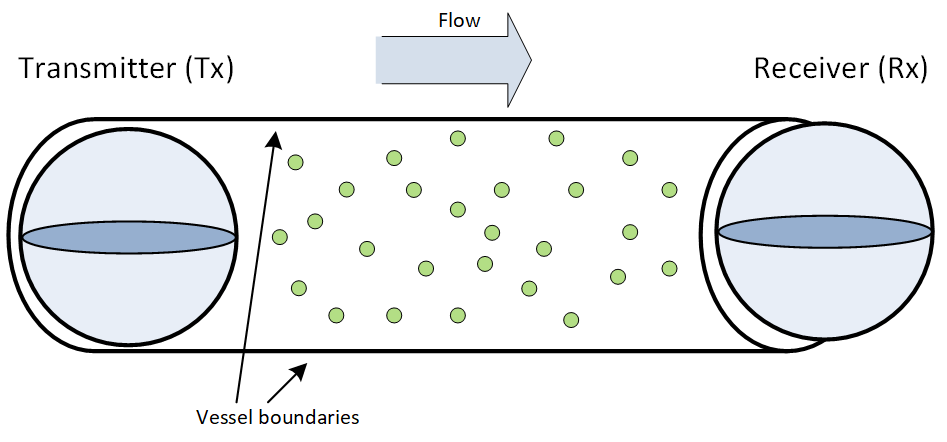}
  \caption{Vessel-like environment with spherical transmitter (Tx) and spherical receiver (Rx).}
  \label{fig:mol_com_vessel_like}
\end{figure}

One commonly-used special case of the constrained diffusion environment is a vessel-like environment that emulates the inside of blood vessels. In these environments, the Tx and Rx pair is considered to be within a cylindrical boundary, which acts as the vessel boundary (Fig.~\ref{fig:mol_com_vessel_like}). In studies focusing on vessel-like environments, the vessel boundary is typically modeled as a reflective surface for the MMs, while the Tx and Rx are located on different sides of the vessel, and a flow is existent in the direction of the Rx. Since generally the flow in blood vessels is laminar \cite{Back1986_Measurement}, studies focusing on vessel-like environments consider laminar flow in their analysis. Due to the symmetrical nature of the vessel-like environment, analytical closed-form solutions are tractable for some of these models. For instance, in \cite{dincc2019generalAA}, Dinc \etal{} developed a general approximation for flow in 3D microfluidic channels. 

\subsection{Channel Characteristics}
\label{subsec:mc_channel_characteristics}

As the MCvD channel is governed by the diffusion process, it becomes a linear system provided that the diffusion coefficient $D$ does not depend on time, space, or molecular density, and that the detection process at the Rx does not alter the linearity of the system \cite{Jamali2018_Channel, Farsad2014_Channel}.

\subsubsection{Channel Impulse Response}

The channel impulse response (CIR) function, which characterizes the channel response to an impulse input, can be used to define a communication channel. 
There are a variety of definitions in the MC literature for the CIR function (i.e., $h(t)$ or $h(t,\tau)$) \cite{ahmadzadeh2017statisticalAO,wicke2018modelingDF,jamali2017scwCodes}. We follow the approach of~\cite{Wang2014_Transmit,luo2019nonCS,Yilmaz2014_3DChannel} and  define the CIR function through another function called the Cumulative Fraction of Received Response (CFRR). Please note that it is also possible to define CFRR after defining CIR, the important point is that CIR is the derivative of CFRR with respect to time (i.e., CFRR is the integration of CIR with respect to time). 
\begin{defn}
The \textit{Cumulative Fraction of Received Response (CFRR)} is defined as the mean fraction of molecules that are detected by the receiver {\em on average} until time $t$ when molecules are emitted at $t=0$. The CFRR is denoted by $F(t)$.
\end{defn}

After defining the CFRR, we can now define the CIR using the CFRR as follows:
\begin{defn}
\textit{Channel Impulse Response (CIR)} is defined as the derivative of CFRR with respect to $t$ and denoted by
\begin{align}
    h(t) = \frac{d}{dt} F(t).
\end{align}
\end{defn}
The CIR physically represents the rate of reception of MMs at the Rx as a function of time. The CIR and the CFRR define important aspects of the molecular channel characteristics. For instance, for an absorbing receiver in a 1D free diffusion environment, with no chemical reaction or decay, CFRR is analytically known and shows that the emitted molecules eventually arrive at the receiver (i.e., ${\lim_{t\rightarrow\infty} F(t) = 1}$). On the other hand, in a 3D free diffusion environment, there is a non-zero probability that a given molecule will never reach the Rx as time goes to infinity (i.e., $\lim_{t\rightarrow\infty} F(t) < 1$). As shown in ~\cite{Eckford2008_Molecular,Srinivas2012_Molecular,Yilmaz2014_3DChannel}, CIR has a heavy tailed structure in 1D-3D environments, which implies that ISI is one of the common reasons for communication errors in an MCvD system. In particular, ISI is introduced by stray MMs belonging to the current time slot that can interfere and impair the detection of future symbols. This issue is exacerbated as more and more stray MMs from all previous time slots accumulate.

In the MC literature, CFRR has been investigated for different channel types such as free diffusion, constrained diffusion, and vessel-like environments. In recent years, closed-form solutions of CFRR have been derived analytically for some key MC channels such as 1D free diffusion (with or without molecular degradation\footnote{Molecular degradation can be defined as the alteration of the chemical structure of the MMs as a result of a chemical reaction. Sometimes this reaction can be naturally occurring and sometimes it can be accelerated by adding enzymes to the environment. For MC, this effect may improve the received signal quality by eliminating the remaining interference molecules or it may deteriorate the received signal by eliminating the MMs too much.}) \cite{ Noel2016_Channel, Nakano2012_Channel}, 3D free diffusion (with or without molecular degradation) \cite{Yilmaz2014_3DChannel, akkaya2015effectOR, heren2015effectOD}, 3D constrained diffusion \cite{Jamali2018_Channel}, and vessel-like environments \cite{Jamali2018_Channel,Turan2018_Transmitter}. However, for more complex channels (e.g., constrained diffusion environment, time-variant diffusion coefficients), finding analytical solutions for CFRR is much more challenging. Hence, for such environments, instead of closed-form solutions, infinite sum formulations have been derived for given environmental parameters, such as the CFRR formulation given for 3D diffusion channels with a reflective spherical source and spherical receiver in \cite{Genc2018_Reception} and a 3D vessel-like environment with drift given in \cite{wicke2018molecular}.

\subsubsection{Channel Capacity}

Another fundamental characteristic of the MC channel is its channel capacity, which provides a fundamental limit for the maximum information transmission rate that is achievable over the MC channel. Note that this capacity is independent of the modulation and coding techniques being used.

As explained in Section~\ref{subsec:mc_channel_properties}, due to the diffusion dynamics, the MC channel is a channel with memory and should be modeled as such \cite{Hsieh2013_Robust, Genc2016_ISI}. Therefore, finding analytical expressions for the channel capacity is challenging \cite{gohari2016_ITforMC,Rose2019_CapacityIEEEProc}. Some prior works have relied on a simplifying assumption that the MC channel is memoryless and evaluated channel capacity or bounds under this assumption. For example, in ~\cite{Einolghozati2011_Capacity,Nakano2012_Channel,Srinivas2012_Molecular,Li2014_Capacity,Farsad2016_CapacityOD,Rose2016_InscribedPartI,Aminian2015_Capacity,Farsad17_PICcapacity,Farsad2019_CapacityLimits}, upper and lower bounds on channel capacity are derived for different memoryless MCvD channels. Some other works have relied on achievable information rate as a measure to compare different modulation techniques. These works will be discussed in more detail in the next sections. 

For the stationary channels, a third approach in channel modeling regarding memory is to use a channel model with a finite memory $m$. These kinds of models only consider the effects of the previous $m$ symbols over the current one. In the MC literature, studies that consider a channel with finite memory of this nature set $m=1$. Here it should be noted that the selection of an adequate $m$ value depends on the symbol duration. Considering the symbol duration values that are common in the MC literature, it has been shown by Genc \etal{} that setting $m=1$ is far from adequate to accurately model channel memory \cite{Genc2016_ISI}. In particular, Genc \etal{} show that $m$ should be on the order of tens of symbols. 

\section{Classification of Modulation Techniques}

Similar to classical communication systems, the modulation process in an MCvD system varies one or more physical properties of the MM signal to transmit bits of information. Since the MM signal is physically different from a traditional EM signal, the physical properties that can be used to modulate bits onto signals in an MCvD system differs from the EM signal properties that are used to modulate bits in classical communication systems, i.e., amplitude, frequency, and phase. 

We now describe and classify the multitude of modulation techniques that have been proposed for MCvD systems. These techniques are classified according to the physical property being used to encode bits into the transmitted symbols: \textbf{concentration}, \textbf{type}, \textbf{timing}, and \textbf{space} modulation. Among these physical properties, concentration refers to the amount of MMs transmitted by the Tx; type refers to the type of MMs transmitted by the Tx; timing refers to a property of the transmission time of MMs within the communication time slot; and space refers to the spatially separated multiple MM release modules at the Tx in a molecular MIMO scenario. Also, some techniques utilize a combination of these physical properties as part of a single modulation technique. We classify these techniques as \textbf{hybrid} techniques.

We divide the characteristics and assumptions of individual modulation techniques into two main categories: modulation characteristics and performance evaluation assumptions. In modulation characteristics, we consider ISI mitigation, computational complexity (i.e., the complexity of both transmission and reception processes), and the detection type used at the Rx. 

We define four qualitative levels for both the ISI mitigation and computational complexity characteristics: none, low, moderate, or high. Among these two characteristics, the computational complexity also depends on the implementation and the type of Tx and Rx devices. It might be the case that a given technique is much easier to implement for biological type devices than described here. However, a discussion on the implementation of these techniques is outside the scope of this survey. As for the detection type, while most of the demodulation techniques utilize basic thresholding (e.g., concentration is above a threshold or not), others use more complex detection types such as maximum likelihood (ML), which generally yield higher overall performance at a cost of higher computational complexity.

Studies on these specific modulation techniques evaluate their performance based on differing sets of assumptions. We categorize the assumptions used in performance analysis as follows: transmitted signal waveform (in terms of number of MMs released over time), receiver type and the system environment. Note that the proposed modulation techniques are independent from these assumptions and that the performances of the considered modulation techniques may differ based on these features.

\section{\label{sec:Perf_Metrics}Performance metrics for Molecular Communication via Diffusion system} 
In the MC literature, authors use a variety of different performance metrics such as bit error rate (BER), achievable information rate, and interference to total received molecule ratio (ITR) to evaluate the system performance. Although the core concepts of these metrics are general to any communication system, some physical concepts and evaluation methods are specific to the MCvD system. In this section we will enumerate such performance metrics and elaborate on their use in the MC literature. Note that these metrics can be interdependent for a given channel and modulation.

\subsection{Error Rates (BER, SER)}

A common metric in evaluating the performance of a communication system is the BER (or symbol error rate (SER) if a given symbol represents more than a single bit of information). In the MC literature, BER or SER is evaluated given an MC channel and modulation technique, and is a function of signal-to-noise ratio (SNR), transmitted power, symbol duration, and environmental parameters (e.g., communication distance and diffusion coefficient).

In classical wireless communication, the BER is usually plotted against transmitter power or SNR. However, when we look at these parameters in the MC domain, the transmitter power is usually the amount of MMs sent from the Tx for the signal in question. In order to determine the SNR for an MC channel, noise as well as its sources must be defined, as the noise in MC channels has different characteristics than in wireless or wired channels. In this regard, many works in the MC literature consider the source of noise to be the diffusion process, giving rise to the notion of diffusion noise\cite{Guo2016_Channel}. This particular noise is based on the probabilistic nature of the arrival of molecules through the MC channel and have a considerable effect on the performance of the communication system. However, the current state of the MC literature lacks a consistent definition of this diffusion noise, which makes performance comparisons between different works somewhat difficult.

Alternatively, BER can be evaluated against signal-to-interference and noise amplitude ratio (SINAR). In this case, in addition to the noise sources, interference sources must also be defined and evaluated. There are several key interference sources for an MC system such as inter-symbol interference (ISI), co-channel interference (CCI) \cite{Kuran2012_Interference}, and inter-link interference (ILI) \cite{koo2016molecularMIMO,gursoy2018indexMF}. Among these sources, ISI is generally considered in performance evaluations, whereas CCI and ILI are generally only considered in works that study multiple transmitters and MIMO systems, respectively.

\subsection{Achievable Information Rate}

Another commonly used performance metric for a given modulation technique is the achievable information rate. Similar to BER, achievable information rate is also usually evaluated against SNR, transmitted power, symbol duration, and environmental parameters such as communication distance and diffusion coefficient. 

In wireless communication, transmitted power is generally given in decibel-milliwatts (dBm) or watts. As described in \cite{Kuran2010_Energy}, the transmitted power in an MCvD system mainly depends on the energy costs of synthesizing MMs and moving these MMs to the outside of the Tx. In both of these accounts, the key multiplier defining the overall transmitted power is the MM count. Hence, in an MCvD systems, the transmitted power can alternatively be given in MM count or MM count per type. Consequently, the achievable information rate can be defined per energy, per MM, per MM type and per unit time.

As mentioned in Section II.C.2, the MC channel typically exhibits memory. For a channel with memory, the mutual information rate can be calculated as
\begin{align}
    I(S;\hat{S}) = \lim_{n\to\infty}\frac{1}{n}I\left(S[1], ... S[n];\hat{S}[1], ... \hat{S}[n]\right)
\end{align}
for a given modulation technique, where $S[k]$ and $\hat{S}[k]$ represent the transmitted bit value and the demodulated bit value during the $k^{th}$ symbol slot (considering the fact that the communication is conducted in a time-slotted manner), respectively \cite{Pfizer2001_Achievable}. As such, several works have evaluated the achievable information rate in such MC channels with memory. These works have mainly considered the concentration shift keying (CSK) modulation technique where information is modulated by varying levels of emitted MM concentration. In \cite{Nakano2012_Channel, Atakan2013_Optimal}, the achievable information rate for a 1D MC channel is evaluated considering a CSK modulation technique and varying channel memory $m$. In  \cite{Singhal2015_Performance, Liu2015_Channel, Genc2016_ISI}, a 3D MC channel is evaluated again by considering a CSK technique and varying channel memory. As opposed to these works, in \cite{Arjmandi2015_Capacity} the achievable information rate for a 1D MC channel is calculated considering a more sophisticated modulation technique of the molecular concentration shift keying (MCSK), where information is modulated by varying types of emitted MM concentration. Both the CSK and the MCSK modulation techniques are explained in detail in Section~\ref{sec:Mod_Tech}. 

\subsection{Interference to Total Received Molecule Ratio (ITR)}

\begin{figure}[t]
  \centering
    \includegraphics[width=1.0\columnwidth]{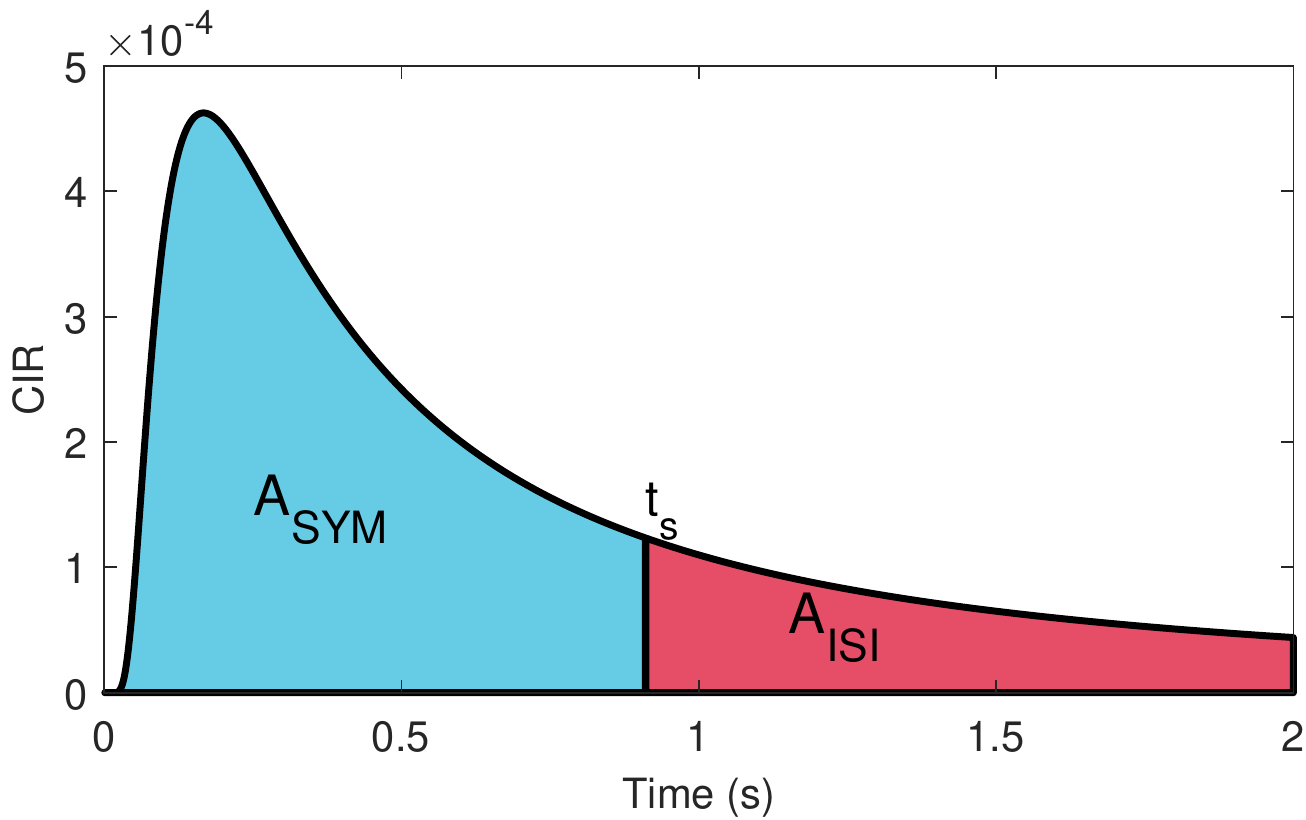}
  \caption{Representation of the molecular received signal and its components: desired signal and ISI. The area $\text{A}_{\text{SYM}}$ corresponds to the cumulative number of received molecules until $t_s$. Molecules inside the interference area ($\text{A}_{\text{ISI}}$) are received after $t_s$ and may cause erroneous detection for upcoming symbols.}
  \label{fig:metrics_itr}
\end{figure}

As previously explained, in most MCvD environments, CIR has a heavy tail structure. The stray MMs from previous symbol slots accumulate and impair the correct reception ability of the Rx. Two components that can be derived from CIR, the area under the desired signal, $A_{\text{SYM}}$, and the ISI, $A_{\text{ISI}}$, both depicted in Fig.~\ref{fig:metrics_itr}. For a fixed symbol duration $t_s$, interference molecules reduce the communication quality by affecting the reception process of the upcoming symbols.

Hence, Interference to Total Received Molecule Ratio (ITR) for a given symbol duration $t_s$ denotes the fraction of interference molecules for a single emission of molecules. It is given by
\begin{align}
    \text{ITR}(t_s) = \frac{A_{\text{ISI}}}{A_{\text{SYM}}+A_{\text{ISI}}}=\frac{\int_{t_s}^\infty h(t) \, dt}{\int_{0}^\infty h(t) \, dt}.
\end{align}
Therefore, ITR for a given $t_s$ is evaluated by considering $h(t)$ or an empirical approximation to it.

Similar to ITR, signal-to-interference and noise amplitude ratio (SINAR), signal-to-interference ratio (SIR) and signal-to-interference difference (SID) can also be used as a molecular signal quality metric. SINAR is analogous to SINR in conventional communications by considering noise and interference molecules. SIR is the fraction of the amount of desired signal molecules to interference molecules and SID is the difference between the number of desired signal and interference molecules. There are various works in the literature that consider ITR~\cite{Mahfuz2011_Characteristics,heren2015effectOD,koo2016molecularMIMO,yilmaz2016interference,cho2017effective}, SINAR~\cite{gursoy2019simulationSA}, SIR~\cite{Tepekule2015_Novel,koo2016molecularMIMO}, and SID~\cite{akdeniz2018molecularSM,akdeniz2018optimalRD} for the performance evaluation. In~\cite{gursoy2019simulationSA}, it is shown that the values of the system parameters that give maximum SID or SINAR values also give close to optimal BER or SER value. Therefore, SID or SINAR can be used as a representative metric for finding close to optimal system parameters in terms of BER or SER~\cite{akdeniz2018molecularSM,akdeniz2018optimalRD,gursoy2019simulationSA} for an MCvD system with similar parameters in the papers. It is observed that in cases where $A_{SYM}$ is not high, but $A_{ISI}$ is very small (i.e., nearly zero), SIR attains extremely high values. Therefore, it is argued that SID is a more reliable performance metric than SIR, for the system parameters that are mentioned in~\cite{akdeniz2018molecularSM,akdeniz2018optimalRD}. However, these important findings need more analysis before generalizing it.

\subsection{Peak-to-Average Messenger Molecule Ratio (PAMR)}

In typical MC systems, for a given symbol the Tx emits all the MMs at once. However, the capabilities of the Tx might prohibit such an emission and instead it might distribute the emission times and the total amount of MM per symbol over the symbol duration.  In such an emission pattern, another metric, called the Peak-to-Average Messenger Molecule Ratio (PAMR) at the Tx, becomes relevant to the performance of the system. PAMR is a similar concept to Peak-to-Average-Power-Ratio (PAPR) in a traditional EM-based communication system. It is defined as
\begin{align}
    \text{PAMR}_{\text{Tx}} = \frac{\max N^{Tx}(t_e)}{\text{avg}\, N^{Tx}(t_e)}, \;\;\; t_e \in \mathscr{T}_e, 
\end{align}
where $\mathscr{T}_e$ is the set of emission times and $N^{Tx}(t_e)$ stands for the number of emitted molecules by the Tx at the emission time $t_e$~\cite{Yilmaz2014_SimulationSO}. Note that in an MC system where the Tx releases all of the MMs at the beginning of a symbol slot, $t_e$ is the time at the beginning of each symbol slot. PAMR is related to transmitter capabilities and implies constraints on the number of emitted molecules over a given time duration. If the Tx has a limited PAMR due to its transmission constraints, it will result in a degradation of the communication performance. 

\subsection{MOL-Eye}
\begin{figure}[t]
  \centering
    \includegraphics[width=1.0\columnwidth]{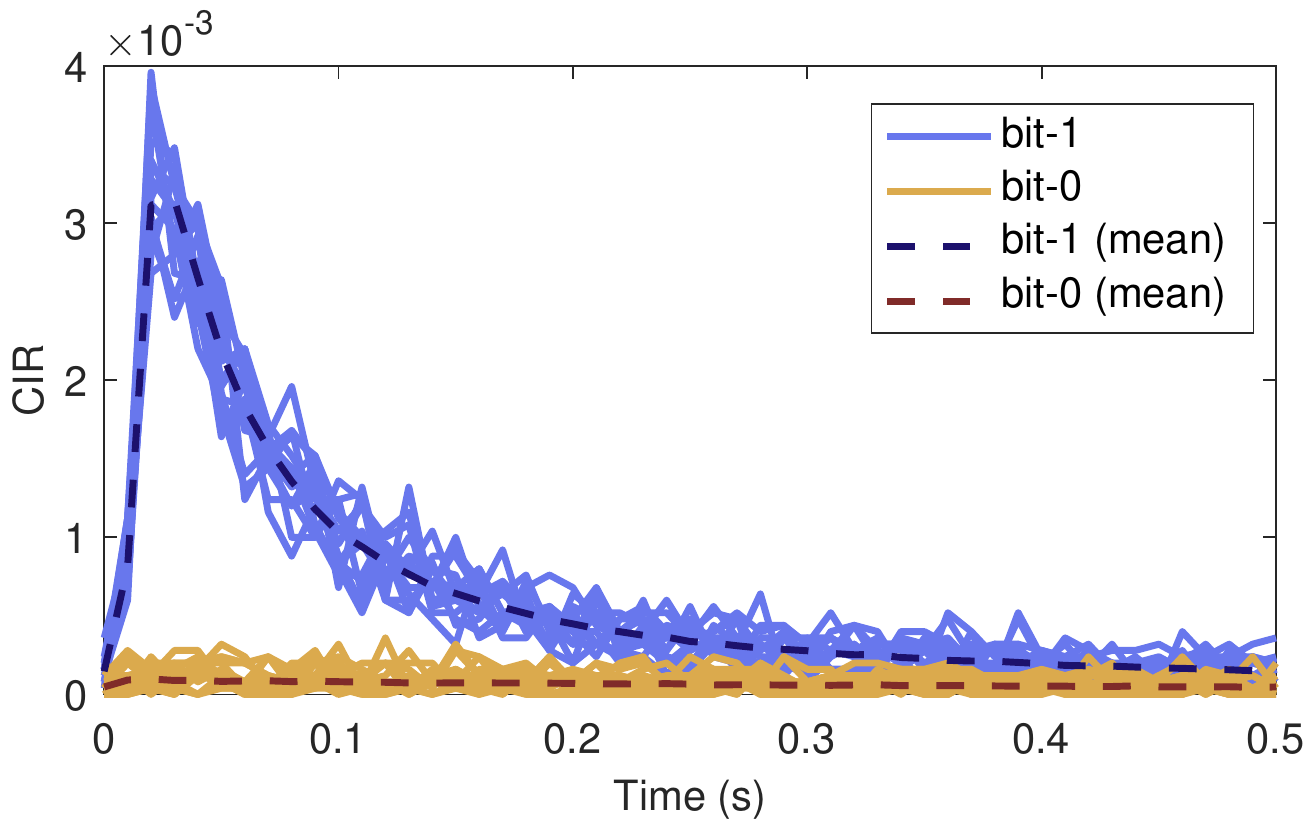}
  \caption{An example MOL-Eye diagram considering a binary system composed of curves of the number of received molecules representing each bit-0 and bit-1 transmissions and their mean curves (adapted from \cite{Turan2018_MOL}).}
  \label{fig:metrics_moleye}
\end{figure}
The eye diagram is another tool used to evaluate the quality of a signal after transmission through a channel  \cite{Grami2016_DigitalComm}. Given a communication channel and  modulation technique, an eye diagram is obtained by superimposing the received signals of consecutive bit transmissions on top of each other. Various features of the resulting diagram can be checked to give a general understanding of the underlying communication system such as when to sample the signal, amount of jitter, and the SNR at the sample point. 
Turan \etal{} propose utilizing a similar tool called the MOL-Eye for the MC domain \cite{Turan2018_MOL} to evaluate the performance of a modulation technique. They propose using three performance metrics associated with the MOL-Eye diagram: the maximum eye height, standard deviation of the received molecules ($STD(\Delta_c)$), and counting SNR (CSNR).

Considering a binary system where each symbol represents a single bit value, curves of the number of received molecules representing each realization of bit-1 and bit-0 transmissions are evaluated and are called  bit-0 and bit-1 curves, respectively (Fig.~\ref{fig:metrics_moleye}). Then, the maximum eye height is calculated as the biggest distance between the mean of bit-0 and mean of bit-1 curves. To evaluate $STD(\Delta_c)$ and CSNR, first the integral difference (that is indicating the area) between every combination of bit-0 and bit-1 curves are calculated as 
\begin{align}
\Delta_c(i,j) =  \int_{0}^{t_s} c_1(i) -  c_0(j) \,dt ,
\label{eq_integralDif}
\end{align}
where $c_1(i)$ and $c_0(j)$ are the $\xth{i}$ bit-1 and $\xth{j}$ bit-0 sampled curves. Using this integral difference, $STD(\Delta_c)$ is calculated as the standard deviation of all possible $\Delta_c(i,j)$ values. Similarly, CSNR is evaluated as
\begin{align}
CSNR = \frac{\mu_{\Delta_c}}{STD(\Delta_c)},
\label{eq_CSNR}
\end{align}
where $\mu_{\Delta_c}$ represents the mean of all possible $\Delta_c(i,j)$ values.

Among these three metrics, it is observed in \cite{Turan2018_MOL} that CSNR has a one-to-one relation to BER. Assuming a channel with finite memory $m$ (i.e., a model that only considers the effect of previous $m$ symbols over the current one), BER is calculated by evaluating all the $2^{m}$ bit sequence combinations. On the other hand, CSNR is evaluated over a fixed number of bit curves (e.g., $100$ in \cite{Turan2018_MOL}). Therefore, it is computationally easier to calculate than BER, hence is used instead of BER to fine tune the operational parameters of devices.

\section{\label{sec:Mod_Tech}Modulation Techniques}
All MC modulation techniques based on releasing MMs from the Tx and modulating the bit-values of a given message onto the features of the emission of MMs. On the Rx end of the system, these molecules interact with special biochemical protein structures called ``receptors" and trigger an event within the cell based on the receptor and the signaling pathway within the Rx\footnote{Signaling pathway is a molecular biology term describing a group of molecules in a cell that work together to control cell functions. After the first molecule in a pathway receives a signal, it activates another molecule. This process is repeated until the last molecule is activated and the cell function is carried out hence a ``pathway" \cite{Alberts2014_MBCell}.}.

There are different modulation techniques proposed in the MC literature. These techniques can be broadly classified into five groups based on the feature of the emission of MMs: 
\begin{itemize}
\item \textbf{Concentration-based Techniques}: Information is represented by the varying concentration level of the transmitted signal.
\item \textbf{Type-based Techniques}: Information is represented by the type of the MMs of the transmitted signal.
\item \textbf{Timing-based Techniques}: Information is represented by various time related features of the transmitted signal.
\item \textbf{Spatial Techniques}: Information is represented by the spatial location of emission, especially for multi-antenna systems.
\item \textbf{Hybrid Techniques}: Techniques utilizing more than one of the four features of the transmitted signal listed above to represent the information.
\end{itemize}
 
\subsection{Concentration-Based Techniques} 

The main idea of concentration-based techniques is carrying information on the released MM concentration over discrete period time slots (i.e., symbol slots), where each slot is used to carry a single symbol of the overall message. In most works, each time slot has a fixed period. In its simplest form where each symbol represents a one-bit value (called on-off keying (OOK)), if the corresponding bit value (\(S[k]\)) is bit-1, the Tx releases a fixed number of MMs (i.e. \(n_1\)). On the other hand, if it is bit-0 the Tx does not release any molecules for that symbol slot~\cite{Mahfuz2010_CharOB}. At the receiver side, the Rx counts the number of MMs that arrive within each symbol slot (i.e., \(N^{Rx}[k]\)) and makes a threshold based decision to decode the bit value of the given symbol slot (\(\hat{S}[k]\)). If \(N^{Rx}[k] \geq \lambda\), \(\hat{S}[k]\) is decoded as bit-1; else it is decoded as bit-0, where \(\lambda\) is a threshold value for signal detection.

\begin{table*}[t]
\begin{center}
\caption{Concentration-Based Modulation Techniques for Molecular Communication }
  \begin{tabular}{ | p{2.3cm} | c | c | c | c | c | c | c |}
    \hline
    \rowcolor{headerColor} 
     \multicolumn{2}{|c|}{\textbf{Technique Identification}}    &  \multicolumn{3}{c|}{\textbf{Modulation Characteristics}}   &  \multicolumn{3}{c|}{\textbf{Performance Evaluation Assumptions}} \\ 
    \hline
    \rowcolor{headerColor}
    Name            & References & ISI Mitig. & Comp. Complex. & Detection Type        & Tx Waveform       & Rx Type   & Environment \\ \hline 
    CSK             & \tabCrefA  & None       & Low        & Threshold          & Pulse           & Absorbing & 3D, No Drift \\ \hline 
    CSK-AD          & \tabCrefC  & None       & Low        & Inst. Threshold    & Pulse           & Passive   & 3D, No Drift \\ \hline
    CSK-SD          & \tabCrefD  & None       & Low        & Threshold          & Pulse           & Passive   & 3D, No Drift \\ \hline
    CSK-SubTS       & \tabCrefE  & None       & Low        & Threshold          & Imperfect Pulse & Absorbing & 1D, with Drift    \\ \hline
    CSK-PA          & \tabCrefF  & High       & High       & Threshold          & Pulse           & Absorbing & 1-2D, No Drift \\ \hline
    CSK-CPA         & \tabCrefFA & High       & Moderate   & Threshold          & Pulse           & Absorbing & 1-2D, No Drift \\ \hline
    CSK w/ ATD      & \tabCrefGA & Low        & Low        & Adaptive Threshold & Pulse           & Absorbing & 3D, No Drift \\ \hline
    CSK w/ ML, MAP, MMSE & \tabCrefG  & Moderate & High       & ML, MAP, MMSE              & Pulse           & Absorbing & 3D, No Drift \\ \hline
    NC-CSK-Diff     & \tabCrefH  & None       & Low        & NC Diff.    & Imperfect Pulse & Passive   & 3D, No Drift \\ \hline
    NC-CSK-Convex   & \tabCrefI  & Moderate   & Low        & NC Convex   & Imperfect Pulse & Passive   & 3D, No Drift \\ \hline
    NC-CSK-Gamma    & \tabCrefJ  & Moderate   & Low        & NC ML       & Pulse           & Passive   & 3D, with Drift    \\ \hline
    CSK w/ Eq. Sig. & \tabCrefK  & High       & Moderate   & Threshold          & Pulse           & Passive & 2D, with Drift    \\
    \hline
  \end{tabular}
  \label{table:CSK-variants}
\end{center}
\end{table*}

A more generalized version of OOK is called Concentration Shift Keying (CSK) where, depending on the system design, each symbol represents \(m\)-bits of information. Following the classical modulation terminology, if a symbol represents one-bit of information, this technique is called Binary CSK (BCSK), and if a symbol represents two-bits of information it is called Quadrature CSK (QCSK), and so on \cite{Kuran2010_Energy, Kuran2011_Modulation, Kuran2012_Interference, Mahfuz2011_Characteristics, Lin2012_Signal}. In CSK, for the \(\xth{k}\) symbol in the message, the Tx releases \(N^{Tx}[k]\) number of MMs depending on the current symbol value as
\begin{equation} \label{eq:CSK_Tx_Mol}
N^{Tx}[k]=n_{S[k]}, \quad S[k] \in \left\{sym_0, sym_1, ..., sym_{2^m-1} \right\} ,
\end{equation}
where $n_{S[k]}$ denotes the number of molecules to be emitted for the symbol value of $S[k]$ and please note that \(S[k]\) can take one of the \(2^m\) symbol-values (e.g., $sym_0$, $sym_1$) in the modulation alphabet. In order to demodulate \(\hat{S}[k]\) from the received signal, the Rx uses \(2^m-1\) thresholds (i.e., \(\lambda_0, \lambda_1, ..., \lambda_{2^m\!-\!2}\)) as

\begin{equation} \label{eq:CSK_Rx_Decode}
\hat{S}[k]= 
  \begin{cases}
    sym_0,  \;\qquad &\!\!\!N^{Rx}[k] < \lambda_0 \\
    sym_i,  \;\qquad\lambda_{i-1} \leq &\!\!\!N^{Rx}[k] < \lambda_{i}, \;\;\;1\!\leq\! i \!\leq\! 2^m\!-\!2\\
    sym_{(2^m\!-\!1)}, \;   \lambda_{2^m\!-\!2} \leq &\!\!\!N^{Rx}[k] \\
  \end{cases}
\end{equation}
where $N^{Rx}[k]$ denotes the number of received molecules during the $\xth{k}$ symbol slot.

From a communication point-of-view, CSK is akin to the Amplitude Modulation (AM) method in analog modulation and Amplitude Shift Keying (ASK) in digital modulation. It inherits the main advantages and disadvantages of ASK as being a simple system to implement, but also considerably sensitive to interference and noise in the environment. As explained in Section~\ref{sec:Classification}, the CIR of the MC channel has a long-tail component. This long-tail component causes considerable ISI, which in turn reduces the correct decoding probability of the threshold based receiver. 

In the literature, numerous CSK-variants have been proposed to increase the communication performance, e.g., to decrease the error probability of the system (Table~\ref{table:CSK-variants}). Most of the CSK-variants consider the Tx releasing all the MMs based on the selected \(N^{Tx}[k]\) value at the start of the corresponding symbol slot. In \tabCrefE, Singhal \etal{} consider sending two dirac pulses during a single symbol slot, one at the beginning and another one in the middle. Correspondingly, Rx utilizes two thresholds, one within the first half of the symbol duration ($k_{fh}$) and another for the second half ($k_{sh}$). It demodulates the received signal as 
\begin{equation} \label{eq:CSK_SubTS_Rx_Decode}
\hat{S}[k]= 
  \begin{cases}
    sym_0, & \qquad\; N^{Rx}[k_{fh}] < \lambda \;\&\; N^{Rx}[k_{sh}] < \lambda\\
    sym_1, & \qquad\;\lambda \leq N^{Rx}[k_{fh}] \;\&\; \lambda \leq N^{Rx}[k_{sh}] \\
    sym_2, & \qquad\qquad\quad \mbox{otherwise} \\
  \end{cases}.
\end{equation}
The resulting technique, called CSK Sub-Timeslot (CSK-SubTS), carries one out of three symbol values per symbol slot. 

In order to minimize the error probability due to ISI, a power adjustment (CSK-PA) technique has been proposed in which the Tx regulates its emission amount (i.e., \(n^{\boldsymbol{S}}_{sym_i}\)), where \(\boldsymbol{S}=(S[k-1], S[k-2],...,S[k-m])\) denotes the values of the past \(m\)-symbols. By this adjustment, CSK-PA aims to minimize the difference between the \(N^{Rx}[k]\) values with the same $S[k]$ values \cite{Einolghozati2011_Capacity, Tepekule2014_Energy, Movahednasab2015_Adaptive}. Considering a BCSK-PA system, the goal is to minimize the variation between the \(N^{Rx}[k]\) values where $S[k] = sym_1$. Since the Tx knows the past \(m\)-symbols, if the channel parameters (i.e, \(D, r_{rx}, d,\)) are also known, this adjustment can be accomplished with near certainty. The calculation of different \(n^{\boldsymbol{S}}_{sym_i}\) values depend on the past \(m\)-symbol values. As \(m\) increases, calculation of the \(n^{\boldsymbol{S}}_{sym_i}\) values becomes impractical in terms of computational power and memory requirements. In all three works focusing on the CSK-PA technique, \(m\) is selected as very small to keep the system simple and practical (In \cite{Einolghozati2011_Capacity}, \(m\) is selected as 1 (i.e., 1-bit memory) while in \cite{Tepekule2014_Energy, Movahednasab2015_Adaptive} it is selected as 2).

A variation of the CSK-PA technique, called the consecutive power adjustment (CSK-CPA) technique, only focuses on the worst case scenarios \cite{Turan2018_MOL}. In this technique, while focusing on a binary system the Tx only considers the past $l$-symbols of consecutive $sym_1$ values (i.e., \(k \in \{0,1,..,l\}\)) for power adjustment. As an example, considering a binary system where $\boldsymbol{S} = \{1,1,1,1,0,1,1\}$, $S[k]=sym-1$, and the past six bit-values are used for power adjustment, CSK-PA will consider {1,1,1,1,0,1} while CSK-CPA will only consider {1,1,1,1} for power adjustment. Consequently, CSK-CPA greatly reduces the possible Tx states and \(n^{\boldsymbol{S}}_{sym_1}\) values, while only slightly reducing the benefits of the PA mechanism.

In CSK, the number of thresholds used is one less than the cardinality of the symbol set. Therefore, CSK suffers from not being able to differentiate between $sym_0$ and no communication (i.e., silence) during a symbol duration. Mahfuz \etal{} proposed a silence detection technique (CSK-SD) that adds one more threshold at the Rx to distinguish between these two cases \cite{Mahfuz2013_GeneralizedSB,Mahfuz2015_Comprehensive,Mahfuz2016_Concentration}. Considering a BCSK modulation, BCSK-SD utilizes two thresholds \(\lambda_0, \lambda_1\) for demodulating the received signal as
\begin{equation} \label{eq:BCSK_SD_Rx_Decode}
\hat{S}[k]= 
  \begin{cases}
    Silence, & \qquad\; N^{Rx}[k] < \lambda_0 \\
    sym_0,       & \lambda_0 \leq N^{Rx}[k] < \lambda_1\\
    sym_1,       & \lambda_1 \leq N^{Rx}[k] \\
  \end{cases}.
\end{equation}

In \cite{Llatser2013_Detection}, Llatser \etal{} consider utilizing an instantaneous threshold-based receiver (CSK Amplitude detection, CSK-AD) instead of a classical threshold-based receiver. The main difference of the instantaneous threshold-based receiver is that, instead of accumulating all the MMs received within the same symbol slot and taking a thresholding-based decision over this accumulated value, the Rx counts the MMs received within a much shorter unit time (i.e., $\Delta t$) and applies the thresholding over this much smaller number. Although this method can be much cheaper and simpler in design than the traditional threshold-based technique, it greatly increases the error probability of the system. In particular, Aijaz and Aghvami show that this amplitude detection requires \(10^6\) times more MMs to be sent than a threshold-based receiver to achieve the same error probability \cite{Aijaz2015_Error}.

Another variation of the basic threshold-based receiver, the adaptive threshold detector (CSK w/ATD), proposes changing the threshold values, $\lambda_i$, adaptively at every symbol slot. Considering a binary case, the Rx adaptively changes the $\lambda_0$ value by setting it to the $ N^{Rx}[k-1]$ value \cite{Damrath2016_LowComp}. CSK w/ATD is a fairly simple variation that does not require considerable additional complexity at the Rx. According to \cite{Damrath2016_LowComp}, performance-wise, CSK w/ATD decreases the error probability for short $t_s$ and high $d$ values, where the ISI component is significant. On the other hand, it does not perform well when the information has long sequences of consecutive $sym_1$ or $sym_0$ values.

As in any communication system, more complex receivers can be used to minimize the detection errors due to ISI and noise in the communication channel. In a sense, these solutions are the receiver-based alternatives to the CSK-PA technique mentioned above. While in CSK-PA, the Tx tries to minimize the variation between the \(N^{Rx}[k]\) values by regulating the \(n^{\boldsymbol{S}}_{sym_1}\) values, in these receiver-based techniques the Rx tries to correctly demodulate \(\hat{S}[k]\) by utilizing likelihood functions and equalizers over \(N^{Rx}[k]\) values and the channel state information (CSI). However, since the Rx does not exactly know the \(S[k]\) values, these techniques are susceptible to error propagation.

In \cite{Kilinc2013_Receiver}, Kilinc and Akan evaluate the performance of four receivers designed to mitigate ISI in a time-varying diffusion environment: sequence based ML, linear equalization based on minimum mean-square error (MMSE),  Maximum a posteriori (MAP), and decision-feedback equalization (DFE). Although these receivers outperform basic threshold-based receivers in terms of communication performance, they utilize complicated matrix or polynomial operations, which greatly increase their computational complexity \cite{Li2016_LComplexity}. Hence, their suitability to nanomachines, which are expected to have low computational capabilities, may be limited.

It has been argued that in an MCvD system, the CSI may be time-varying or very costly to be exactly acquired by the Rx at each symbol slot. Considering such cases, several works propose receivers that work without having perfect CSI information, which are called non-coherent detectors in MC domain. These detectors/receivers utilize the general  characteristics of the received signal regardless of the values of the environmental parameters (i.e, \(D, r_{rx}, d\)). In \cite{Li2016_LComplexity}, a receiver that utilizes the peak time of the molecular signal and does the demodulation using the difference of the accumulated concentration in two consecutive symbol slots is proposed (i.e., NC-CSK-Diff). Here the peak time of the molecular signal refers to the time value where the maximum number of molecules are received. A similar work, \cite{Li2016_LocalC}, exploits the fact that a molecular signal, i.e., $h(t)$ function, has a singular convex region, which is depicted by investigating the second derivative of $h(t)$. Based on this convex region, the receiver can approximate the ISI effect of a given signal over the subsequent symbol slots without knowing the exact CSI of the channel, and use this estimate to correctly decode the signal. A similar approach is proposed in~\cite{zhai2018antiID} and its performance evaluated on a testbed. All of these works are low-complexity solutions, as is needed in practical applications of nanomachines. 

A third work builds an ML-based non-coherent receiver that utilizes the fact that for a point Tx, spherical passive Rx and an environment with destroyer MMs, the CIR can be approximated by a Gamma distribution \cite{Jamali2016_NonCoherent}. By observing the channel's behavior for several symbol slots, the Rx can estimate the distribution parameters and use them to correctly demodulate \(\hat{S}[k]\) from the signal. The authors propose two versions of the same receiver, one having higher performance and also higher computational complexity, another having very low computational complexity at a cost of worse performance.

Another direction for the receiver design in concentration-based techniques is to take the geometries of the environment, the Tx, and the Rx into consideration in the reception process. In \cite{Akdeniz2020_Equilibrium}, such a technique, named CSK w/Eq. Sig., the authors design a reception process that combines the thresholding and low memory symbol-by-symbol detection in the equilibrium state of the diffusion channel. Although it is robust to variations in the diffusion coefficient as well as variations in the overall geometry of the system components, since it requires the system to reach the equilibrium state, the symbol duration is very high and consequently the technique is suitable for applications with very low bit rate requirements.

\subsection{Type Based Techniques}

\begin{table*}[t]
\begin{center}
\caption{Type-Based Techniques for Molecular Communication}
  \begin{tabular}{ | c | c | c | c | c | c | c | c |}
    \hline 
    \rowcolor{headerColor}
     \multicolumn{2}{|c|}{\textbf{Technique Identification}}    &  \multicolumn{3}{c|}{\textbf{Modulation Characteristics}}   &  \multicolumn{3}{c|}{\textbf{Performance Evaluation Assumptions}} \\ 
    \hline
    \rowcolor{headerColor}
    Name           & References & ISI Mitig. & Comp. Complex. & Detection Type      & Tx Waveform  & Rx Type   & Environment \\ \hline 
    MoSK           & \tabTrefA  & Moderate   & Moderate       & Threshold        & Pulse       & Absorbing & 3D, No Drift \\ \hline
    IMoSK          & \tabTrefB  & Moderate   & Moderate       & Threshold        & Pulse       & Absorbing & 1D, Drift \\ \hline
    MoSK w/ ML     & \tabTrefC  & High       & High           & ML                  & Pulse       & Passive   & 2D, No Drift \\ \hline
    RSK            & \tabTrefD  & Moderate   & Moderate       & Threshold        & Pulse       & Absorbing & 1D, Drift    \\ \hline
    MCSK           & \tabTrefE  & High       & Moderate       & Threshold        & Pulse       & N/A       & N/A  \\ \hline
    MTSK           & \tabTrefF  & High       & High           & Threshold        & Pulse       & Absorbing & 2D, No Drift \\ \hline
    Pre-eq. CSK    & \tabTrefG  & High       & High           & Threshold (Diff) & Pulse       & Absorbing & 3D, No Drift \\ \hline
    Zebra-CSK      & \tabTrefH  & High       & High           & Threshold        & Pulse       & Absorbing & 3D, No Drift \\ 

    \hline
  \end{tabular}
  \label{table:MoSK-variants}
\end{center}
\end{table*}
A second group of modulation techniques, called type-based techniques, focuses on using multiple types of MMs in the communication system as the basis of the modulation technique. In these techniques, the transmitter has the capability of releasing different types of MMs (\(mm_{\text{type}}\)), which are similar to each other in composition (i.e., radius, diffusivity) but can only be received by a particular type of receptor at the receiver surface. Using different types of receptors each corresponding to a particular type of MM, the receiver is capable of receiving multiple molecular release signals, which are practically orthogonal to each other, within a single symbol slot. The quantitative feature of these orthogonal molecular release signals used to represent bit-values depend on the particular technique in question.

In the first type-based modulation technique, Molecular Shift Keying (MoSK), each different symbol-value (\(S[k]\)) is represented by a specific type of MM \cite{Kuran2011_Modulation, Kuran2012_Interference, Aminian2015_Capacity, Galmes2016_Performance}. MoSK uses two types of MMs to modulate one-bit of information within a symbol (called Binary MoSK - BMoSK) or four types of MMs to modulate two-bits of information within a symbol (called Quadruple MoSK - QMoSK). Considering BMoSK, for each symbol slot the Rx counts the number of arriving MMs for each MM type and demodulates \(\hat{S}[k]\) based on the thresholding decisions for each MM type as:
\begin{equation} \label{eq:BMSK_Rx_Decode}
\hat{S}[k]= 
  \begin{cases}
    sym_0,       & N^{Rx}_{mm_{a}}[k] \geq \lambda \;\&\; N^{Rx}_{mm_{b}}[k] < \lambda \\
    sym_1,       & N^{Rx}_{mm_{b}}[k] \geq \lambda \;\&\; N^{Rx}_{mm_{a}}[k] < \lambda \\
    e,       & \mbox{otherwise}
  \end{cases},
\end{equation}
where $a$ and $b$ represent the types of the MMs used, \(MM_t \in (mm_a, mm_b)\), and \(N^{Rx}_{mm_{\text{type}}}[k]\) represents the number of \(mm_\text{type}\) received at the Rx during the \(k^{th}\) symbol slot.

An alternative approach in type-based techniques is to use a majority based detection instead of thresholding \cite{Tepekule2015_Novel}. In such a binary type-based technique, \(\hat{S}[k]\) is decoded as 
\begin{equation} \label{eq:BMSK_Rx_Decode_Majority}
\hat{S}[k]= 
  \begin{cases}
    sym_0,       & N^{Rx}_{mm_{a}}[k] >  N^{Rx}_{mm_{b}}[k] \\
    sym_1,       & N^{Rx}_{mm_{b}}[k] \geq N^{Rx}_{mm_{a}}[k]
  \end{cases}.
\end{equation}

A more generalized version of MoSK has been proposed by removing the constraint of a single type emission during a symbol slot named generalized MoSK (GMoSK) \cite{Chen2020_Generalized}. By utilizing multiple types of MMs, the cardinality of the symbol set is increased which results in higher data rates at higher transmission powers at the cost of higher receptor complexity than the MoSK.

Compared to CSK, MoSK considerably reduces the ISI element of the signal due to the fact that each bit value is represented by a different type of molecule. On the other hand, it increases the complexity of both the Tx and the Rx since the Tx is required to synthesize different types of molecules and the Rx is required to have multiple receptors on its surface. Moreover, the average number of emitted molecules is higher in MoSK compared to CSK since there is an emission of molecules at each symbol slot, while there are symbol slots with no emission in CSK such as while transmitting sym-0 values in BCSK. Unlike CSK, MoSK does not have a direct counterpart among modulation techniques utilized in classical telecommunication. Since each molecule type is practically orthogonal to each other, one can think of them as signals of different frequencies with no overlapping components but the properties of RF signals and molecular signals are widely different. Beside the basic MoSK, other type-based techniques have also been proposed in the literature each focusing on different methods of utilizing multiple MM types in representing the received molecular signal (Table~\ref{table:MoSK-variants}).

As an alternative to the simple thresholding-based detector proposed for MoSK receiver, ShahMohammadian \etal{} propose an ML detector for MoSK in a linear, time-invariant diffusion environment with both ISI and noise components in \cite{ShahM2012_Optimum}. Similar to the work by Kilinc and Akan in \cite{Kilinc2013_Receiver} for concentration-based techniques, although this ML receiver outperforms the simple thresholding-based detector, it requires considerable computation power at the receiver due to the Viterbi algorithm used for calculations, which reduces its practicality inside a nanomachine.

In \cite{Kim2013_Novel}, Kim and Chae propose a type-based modulation technique called ratio shift keying (RSK) using two types of MMs \((mm_a, mm_b)\) and the information is modulated over the received ratio of these two MM types during each symbol duration (i.e., \(R^{Rx}_{mm_{a/b}}[k] = N^{Rx}_{mm_{a}}[k]/N^{Rx}_{mm_{b}}[k] \)). In Binary RSK (BRSK), the received signal is demodulated by using a single threshold value \(\lambda\) as
\begin{equation} \label{eq:BRSK_Rx_Decode}
\hat{S}[k]= 
  \begin{cases}
    sym_0,     & R^{Rx}_{mm_{a/b}}[k] < \lambda \\
    sym_1,     & R^{Rx}_{mm_{a/b}}[k] \geq \lambda \\
  \end{cases},
\end{equation}
whereas in Quadruple RSK (QRSK) it is decoded by using three threshold values \(\lambda_0, \lambda_1\, \lambda_2\) as
\begin{equation} \label{eq:QRSK_Rx_Decode}
\hat{S}[k]= 
  \begin{cases}
    sym_0,       & \qquad\; R^{Rx}_{mm_{a/b}}[k] < \lambda_0 \\
    sym_1,       & \lambda_0 \leq R^{Rx}_{mm_{a/b}}[k] < \lambda_1 \\
    sym_2,       & \lambda_1 \leq R^{Rx}_{mm_{a/b}}[k] < \lambda_2 \\
    sym_3,       & \lambda_2 \leq R^{Rx}_{mm_{a/b}}[k]  \\
  \end{cases}.
\end{equation}
As shown in their work, RSK has a similar performance to MoSK in the binary case but its performance degrades to the level of CSK in the quadruple case due to the fact that it utilizes multiple threshold values, which increases the reception error.

Arjmandi \etal{} propose an alternative type-based modulation technique, called Molecular Concentration Shift Keying (MCSK), where, if \(S[k]=sym_1\), the Tx releases \((mm_a)\) type of MMs for an odd-numbered symbol and releases \((mm_b)\) type of MMs for an even-numbered symbol \cite{Arjmandi2013_Diffusion}. Similar to BCSK, in case \(S[k]=sym_0\) the Tx does not release any MMs. The Rx demodulates the signal by checking the concentration of \((mm_a)\) or \((mm_b)\) depending on the symbol parity as
\begin{equation} \label{eq:BMCSK_Rx_Decode}
\hat{S}[k]= 
  \begin{cases}
    sym_0,       & 
    	\begin{cases}
    	k \;mod \;2 = 0  & N^{Rx}_{mm_{a}}[k] < \lambda \\
        k \;mod \;2 = 1  & N^{Rx}_{mm_{b}}[k] < \lambda \\
  		\end{cases} \\
    sym_1,       & 
    \begin{cases}
    	k \;mod \;2 = 0  & N^{Rx}_{mm_{a}}[k] \geq \lambda \\
        k \;mod \;2 = 1  & N^{Rx}_{mm_{b}}[k] \geq \lambda \\
  		\end{cases} \\
  \end{cases}.
\end{equation}

According to the results provided in \cite{Arjmandi2013_Diffusion}, in the binary case, MCSK outperforms MoSK in terms of BER due to the fact that the ISI component of the molecular signal after a sequence of $sym_1$-valued symbols grows large. Hence it is highly probable that the next symbol with $sym_0$ will be decoded incorrectly as $sym_1$. The alternating approach of MCSK remedies such erroneous decodings. However as shown in \cite{Arjmandi2013_Diffusion}, MCSK loses its edge over MoSK in terms of BER in the quadruple case. So, its advantage is more specific to the binary implementations.

In a communication system, interference can be either constructive or destructive. Considering the BMoSK technique, the ISI part of a $sym_1$ transmission becomes a constructive interference source for the upcoming $sym_1$ symbols (i.e., for all $S[k+x]=sym_1$ where $x \in \mathbb{Z}^+$). On the other hand, for the symbol values $sym_0$ after a $sym_1$ (i.e., for all $S[k+x]=sym_0$ where $x \in \mathbb{Z}^+$, if $S[k]=sym_1$), the same ISI becomes destructive. Tepekule \etal{} propose a technique called molecular transition shift keying (MTSK), which focuses on constructive and destructive ISI differentiation by using two types of MMs \cite{Tepekule2014_Energy}. In MTSK, if $S[k]=sym_1$, the transmitter checks the value of \(S[k+1]\) and chooses one of the two types of MMs according to this value to release in this symbol duration as 
\begin{equation} \label{eq:BMTSK_Tx_Encode}
T_{mm}[k]= 
  \begin{cases}
    mm_{a},       & S[k] = sym_1 \;\&\; S[k+1] = sym_1 \\
    mm_{b},       & S[k] = sym_1 \;\&\; S[k+1] = sym_0 \\
    \;\;\emptyset\;,    & S[k] = sym_0 \\
  \end{cases}.
\end{equation}
Similar to BCSK, in case \(S[k]=sym_0\) the Tx does not release any MMs. At the Rx, the signal is demodulated by checking the concentration levels of both \((mm_a)\) and \((mm_b)\) each symbol slot as
\begin{equation} \label{eq:BMTSK_Rx_Decode}
\hat{S}[k]= 
  \begin{cases}
    sym_0,       & N^{Rx}_{mm_{a}}[k] < \lambda \;\&\; N^{Rx}_{mm_{b}}[k] < \lambda \\
    sym_1,       & N^{Rx}_{mm_{a}}[k] \geq \lambda \;\;|\;\; N^{Rx}_{mm_{b}}[k] \geq \lambda \\
  \end{cases}.
\end{equation}

This way, the Tx utilizes the advantage of the constructive ISI component while at the same time reducing the detrimental effect of the destructive ISI by switching to the second type of molecules when the symbol transition occurs from $sym_1$ to $sym_0$. As expected, MTSK considerably reduces the BER in the system compared to both basic CSK and MoSK techniques, while only increasing the system complexity slightly.

Beside the communication aspects of the type-based techniques, MMs must be selected carefully to be chemically similar to each other and to have high bio-compatibility, especially for the {\em in-vivo} environments. Also, each MM type should have more or less the same diffusion capabilities (i.e., diffusion coefficient) to eliminate any inequality between different symbol-values. In \cite{Kuran2011_Modulation}, hydrofluorocarbon-based MMs are proposed, which are chemically similar to each other and can be easily synthesized by a biological nanomachine. Alternatively, Kim and Chae propose using aldohexoses-based isomers as MMs, called Isomer Molecular Shift Keying (IMoSK), which have much higher bio-compatibility than the hydrofluorocarbons and have similar chemical and synthesis features to them \cite{Kim2012_Novel, Kim2014_SymbolIO}. However, these works are very preliminary studies of MM type selection and none of them shows the bio-compatibility of these molecules in a quantitative fashion. Additionally, these MMs may not be suitable in an in-vitro environment with completely different environmental aspects.

In \cite{Pudasaini2014_Robust}, Pudasini \etal{} propose an advanced version of the MCSK technique called Zebra CSK, where destroyer molecules (called inhibitors) are used to reduce the ISI effect of the received signal. In Zebra CSK, when \(S[k]=sym_1\), in addition to alternatively releasing \((mm_a)\) or \((mm_b)\) depending on the symbol number as in MCSK, the Tx also releases destroyer molecules of type \((dm_b)\) when it releases \((mm_a)\), or \((dm_a)\) when it releases \((mm_b)\). These destroyer molecules interact with the MMs in the environment and practically ``destroy" them in the environment. Therefore, the effect of MMs from previous symbol slots are reduced to have a lower BER.
 
In a similar work, Tepekule \etal{} approach the problem of ISI mitigation by utilizing two types of molecules, \((mm_a)\) and \((mm_b)\), where the Rx measures the received concentrations of each MM type separately and consider the difference of these two concentrations as the actual received signal \cite{Tepekule2015_Novel}. This pre-equalization technique, Pre-eq CSK, works similar to the BCSK technique using the \((mm_a)\) type of MMs. However, when \(S[k]=sym_1\), after releasing type \((mm_a)\) molecules, the Tx also releases type \((mm_b)\) MMs after a short delay. The receiver demodulates the signal by checking the concentration difference between these two types of MMs as
\begin{equation} \label{eq:BPreEqCSK_Rx_Decode}
\hat{S}[k]= 
  \begin{cases}
    sym_0,       & N^{Rx}_{mm_{a}}[k]-N^{Rx}_{mm_{b}}[k] < \lambda \\
    sym_1,       & N^{Rx}_{mm_{a}}[k]-N^{Rx}_{mm_{b}}[k] \geq \lambda \\
  \end{cases}.
\end{equation}

This short delay choice and the ratio between $N^{Rx}_{mm_{a}}$ and $N^{Rx}_{mm_{b}}$ are tunable parameters whose ideal values can be found via heuristic methods depending on the environment. Pre-eq CSK is a less-complex technique from an implementation point-of-view and it achieves a much lower BER than CSK, CSK with MMSE, and MoSK.

\subsection{Timing Based Techniques} 

The third group of modulation techniques encodes information on the time of release of MM. Here, we will refer to this subclass of MC channels as molecular timing channels (MTC). This modulation is different from the previous methods in that the channel input is fundamentally continuous instead of discrete. The MTC model was first proposed in \cite{Eckford2007_Nanoscale}, which is more suitable for evaluating timing-based modulations. In its simplest form, a MTC is based on a single MM released by the Tx at time $t_r$ with information encoded on this release time (RT). The MM goes through some random propagation and arrive at the destination at time
\begin{align}
\label{eq:singleMM_MTC}
 t_y = t_r + t_n,   
\end{align}
where $t_n$ is some random delay due to MM propagation. Note that unlike previous modulations where the symbol set is finite, in pure timing channels, the symbol set is a continuous interval.

An important parameter of the MTC is the probability distribution of the random delay $t_n$. Considering a 1D environment with an absorbing receiver, this random delay is L\'evy distributed for free diffusion and inverse Gaussian distributed for diffusion with drift \cite{Li2014_Capacity}. In a 3D free diffusion environment, as described in subsection \ref{subsec:mc_channel_characteristics}, the MM may never arrive at the Rx, and therefore the channel model must be modified to reflect this effect, as we will show later. A 3D vessel-like environment with drift can be approximated as a 1D environment \cite{Chahibi2013_Molecular} so the same distributions can be utilized for such environments. It can be argued that this probability distribution of $t_n$ has a connection with the CIR of the corresponding MC channel. Particularly, if we assume that the MMs move independent of each other, the expected number of MMs that arrive at the Rx as a function of time (i.e., $E[N^{Rx}[t]]$) is proportional to this probability distribution.  

\begin{figure*}[t]
  \centering
    \includegraphics[width=1.6\columnwidth]{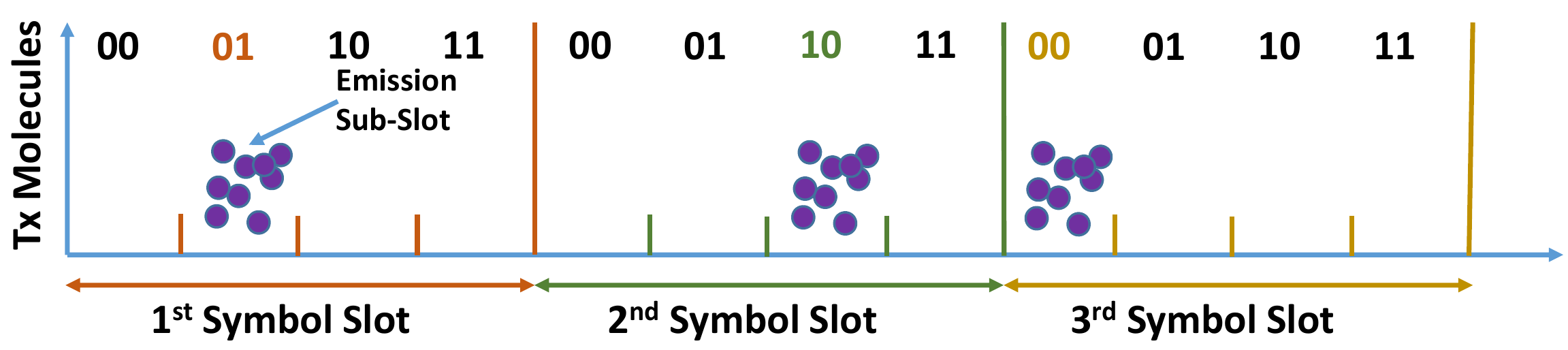}
  \caption{Example scenario of 4-PPM modulation. The information symbols are 01 10 00 and the emissions are done during the corresponding sub-slots. }
  \label{fig:mc_ppm_concept}
\end{figure*}

The first series of works that explore timing-based techniques focus on understanding the fundamental limits of MTCs. In \cite{Srinivas2012_Molecular}, an MTC with additive inverse Gaussian noise (AIGN) was introduced where $t_n$ in \eqref{eq:singleMM_MTC} is inverse Gaussian distributed. The authors derive the upper and lower bounds on capacity per channel use\footnote{A channel use is a single use of the channel to transmit one symbol of the symbol set from the Tx to Rx.} for this MTC with AIGN case. Then, in \cite{Li2014_Capacity}, tighter bounds on capacity of the same MTC with AIGN case were derived. The capacity-achieving MM input distribution was also characterized for this case.

One of the drawbacks of the model in \eqref{eq:singleMM_MTC} is that it only considers a single MM release. In practical systems, many MMs can be released by the Tx. Another drawback is that the notion of channel use (i.e., how long is the channel use duration) is not rigorously defined. It is also assumed that MMs will eventually arrive at the Rx and they are not degraded or destroyed prior to arrival. Finally, ISI is not considered in this model. To overcome some of these limitations, two approaches have been considered in the literature. 

\begin{table*}[t]
\begin{center}
\caption{Timing-Based Techniques for Molecular Communication}
  \begin{tabular}{ | c | c | c | c | c | c | c | c |}
    \hline 
    \rowcolor{headerColor}
     \multicolumn{2}{|c|}{\textbf{Technique Identification}}    &  \multicolumn{3}{c|}{\textbf{Modulation Characteristics}}   &  \multicolumn{3}{c|}{\textbf{Performance Evaluation Assumptions}} \\ 
    \hline
    \rowcolor{headerColor}
    Name               & References   & ISI Mitig. & Comp. Complex. & Detection Type                & Numb. MM  & Rx Type   & Environment \\ \hline 
    RT-Single          & \tabTimerefA & None       & Medium         & ML                            & Single    & Absorbing & 1D, Drift \\ \hline
    RT-Multi           & \tabTimerefB & None       & High           & N/A                           & Multi     & Absorbing & 1D, No Drift \\ \hline
    RT-Multi w/Viterbi & \tabTimerefF & None       & High           & ML/Viterbi                    & Multi     & Absorbing & 1D, No Drift \\ \hline
    PPM                & \tabTimerefD & Low        & Low            & Threshold                 & Multi     & Absorbing & 3D, No drift    \\ \hline
    PPM w/ ML          & \tabTimerefE & None       & Medium         & ML/Max                        & Multi     & Absorbing & 1-3D, Drift  \\ \hline
    PPM w/ decay       & \tabTimerefC & None       & Medium         & First Arrvl/Avg. & Multi & Absorbing & 1D, No Drift \\ \hline
    Time btwn pulses   & \tabTimerefG & Low        & Medium         & Threshold    & Multi/Single & Absorbing & 1-3D, Drift \\ \hline
    MFSK               & \tabTimerefH & None       & High           & Bandpass       & Multi     & Absorbing & 1-3D, No Drift \\ \hline
  \end{tabular}
  \label{table:FSK-variants}
\end{center}
\end{table*}

First, in \cite{Eckford2007_Nanoscale,Eckford2008_Molecular,Rose2016_InscribedPartI,Rose2016_InscribedPartII} the single particle release is extended to multiple MM releases. In this scenario, information is encoded in the vector $\vec{t}_s$, where each element in the vector is the time of release of one of $M$ MMs. It is assumed that each particle goes through a random independent path (hence a random independent delay). The Rx observes the vector $\vec{t}_y$, which contains a set of $M$ arrivals corresponding to each MM. Note that here again it is assumed that all the particles eventually arrive at the Rx. If all $M$ MMs are of different MM types, which do not react or interact with each other, this channel reduces to $M$ parallel channels given in \eqref{eq:singleMM_MTC}. The nontrivial scenario is when the $M$ MMs are of the same type and hence are indistinguishable at the Rx. In this nontrivial scenario, the channel model becomes:
\begin{align}
    \label{eq:multiMM_SortChan}
    \Tilde{\vec{t}}_y = \textsf{sort}\left(\vec{t}_r+\vec{t}_n \right),
\end{align}
where $\vec{t}_n$ is the vector of random delays associated with each MM, $\textsf{sort}(\vec{t})$ is the sort operator that permutes the vector $\vec{t}$ in ascending order of arrival times, and $\Tilde{\vec{t}}_y$ is the observation at the Rx. Note that the sort operation is necessary since the MMs may arrive out of order (i.e., not according to the order they were released) because of independent random delay associated with their random propagation.   

In \cite{Rose2016_InscribedPartI,Rose2016_InscribedPartII}, the fundamental limits of multiple MM release MTC in \eqref{eq:multiMM_SortChan} is investigated. Since this is a complicated channel, it is assumed that the random delays have a finite mean. Particularly, it is assumed that the random delays are exponentially distributed. Then, a guard interval is placed between different channel uses in order to ensure that the probability of ISI is low. Using these assumptions, the $\varepsilon$-capacity\footnote{$\varepsilon$-capacity refers to the maximum  rate of sequence of codes that can attain a decoding error probability less than $\varepsilon\in[0,1)$} of \eqref{eq:multiMM_SortChan} is analyzed by presenting its upper and lower bounds. In $\varepsilon$-capacity, instead of driving the probability of error to zero (i.e., $\varepsilon=0$) as the block length increases, the probability of error is driven down to $\varepsilon,\varepsilon>0$. Note that for $\varepsilon>0$, despite the guard interval, there is a nonzero probability that there will be an ISI event, which is counted as an error in this setup. Note that calculating the capacity of this channel such that the error probability approaches zero is very challenging and that is why the authors consider $\varepsilon$-capacity.

Although \cite{Rose2016_InscribedPartI,Rose2016_InscribedPartII} consider multiple MMs and ISI, as well as rigorously defining the channel use interval, it assumes that all the MMs arrive at the Rx. One of the challenges in MTC channels is that, without this assumption, analyzing its the fundamental limits in \eqref{eq:multiMM_SortChan} becomes very difficult. Therefore, in \cite{Farsad2016_CapacityOD,Farsad2019_CapacityLimits} a new MTC is presented where the MMs have a finite lifetime $\tau_n$. In this channel model, information is encoded in the time of release of $M$ MMs and all of the $M$ MMs are released simultaneously at the same time within a symbol interval $\tau_s$. Consequently, the overall arrival time, $\vec{Y}[i]$, in this MTC channel is given as
\begin{align}
\label{eq:multiMM_DecayMTC}
\vec{Y}[i] = \begin{cases} \vec{t}_{y}[i]= t_{r} + \vec{t}_{n}[i], & \vec{t}_{n}[i] \leq \tau_n \\ \phi, & \vec{t}_{n}[i] > \tau_n \end{cases},
\end{align}
where the $t_{r}$ is the time that the $M$ MMs are released, $\vec{t}_{n}[i]$ is the random propagation delay associated with the \xth{i}  MM, and $\vec{t}_{y}[i]$ is the arrival time of the \xth{i} MM. Here, the first case depicts the case if the MM arrives at the Rx, and the second case corresponds to the case where the MM is degraded before arriving at the Rx. In \cite{Farsad2019_CapacityLimits}, the capacity of the MTC in \eqref{eq:multiMM_DecayMTC} is analyzed by deriving two lower bounds on capacity as well as an upper bound on the capacity. For the case where the random delay is L\'evy distributed (i.e., for free diffusion-based MTC), it is shown that the capacity scales at least polylogarithmically with the number of MMs released. 
 
Besides these information theoretic works, which aim to evaluate the fundamental limits of the MTC, another series of works study system design (e.g., receiver design) for the MTC. Some of these works focus on pulse position modulation (PPM), where information is encoded in the time of release of pulses (i.e., release of a large number of MMs). In this scheme, the symbol interval is divided into $|\mathcal{S}|$ sub-intervals, $t_0, t_1, \cdots, t_{|\mathcal{S}|\!-\!1}$, where $\mathcal{S}$ is the symbol set. Then a pulse of $M$ MMs are released at the beginning of the sub-intervals corresponding to the transmission symbol. Let $y_0, y_1, \cdots, y_{|\mathcal{S}|\!-\!1}$ be the number of MMs that arrive during each sub-interval. Then, assuming that the CIR has its maximum within each sub-interval and that there is no ISI, the symbol can be demodulated using
\begin{align}
    \label{eq:PPM_demodulation}
    \hat{S} = sym_\ell, \;\;\;{\ell = \,\,\margmax\limits_{0 \leq s < |\mathcal{S}|} \, y_s} .
\end{align}
An example scenario of 4-PPM is depicted in Fig.~\ref{fig:mc_ppm_concept}. A summary of various MTC and non-MTC timing-based techniques proposed in the literature are given in (Table~\ref{table:FSK-variants}).

In \cite{Garralda2011_Diffusion}, the performance of CSK and PPM has been compared and it is demonstrated that for binary symbols, CSK is better than PPM in an ideal diffusion scenario. Later in \cite{Akdeniz2018_PPM} it was demonstrated that when the symbol set contains more elements (e.g. 16-PPM), PPM can outperform CSK in terms of BER performance again. This is because PPM is exposed to less ISI. The PPM technique is also investigated in~\cite{zare2019receiverDA} with ML and the proposed (MAX) detectors. The proposed MAX detector records the maximum values in each time bin and detects the intended symbol accordingly.

The optimal ML detector for one-shot PPM (i.e., where there is no ISI), for a detector that can detect the arrival time of individual MMs, is derived in \cite{Murin2019_OneShotPPM}. Since this ML detector can have a large computational complexity, a new detector is proposed that can detect the symbols using the time of arrival of the first MM. It is shown that in some regimes, the performance of this detector approaches to that of the ML detector which can observe the arrival time of all the MMs. These results are extended in \cite{Murin2018_PPMOrderStat}, where it is shown that for a PPM scheme that is similar to the channel in \eqref{eq:multiMM_DecayMTC}, the first arrival receiver, last arrival receiver or the average receiver can each achieve a performance close to the optimal detector, depending on the distribution of $\vec{t}_{n}$. The first arrival detector detects the symbols using the time of arrival of the first MM that arrives at the receiver. Similarly, the last arrival uses the last arrival time of the final MM that arrives at the receiver, while the average arrival time detector can observe the average arrival time of MMs. The optimal detector can observe the arrival time of all the MMs.  

In \cite{Murin2017_SortDetect}, the MTC in \eqref{eq:multiMM_SortChan} is considered, where the Tx can select the release time of individual MMs, and the Rx can detect the arrival time of each MM. The optimal ML detector for such a system is derived and shown to have an exponential computational complexity. Then a sequence detector based on the Viterbi algorithm is derived that achieves a performance close to that of the optimal detector.

One of the drawbacks of encoding information on to the time of release of MMs or PPM is that the Tx and the Rx need to be synchronized. One way to perform asynchronous modulation is to encode information on the time interval between two consecutive releases of molecules or pulses \cite{Garralda2011_Diffusion}. One of the first works to consider asynchronous communication over MTCs was \cite{Hsieh2013_Asynch}, where a system that encodes information on the time interval between molecule releases as well as molecule types was considered. The probability of error and achievable information rates are then derived for the asynchronous scheme and compared with the synchronous schemes.  

In \cite{Krishnaswamy2013_Time-Elapse}, an MC system that relies on bacteria colonies as receivers is considered. Such a system suffers from a large delay since, besides the time it takes for the MMs to travel from the Tx to the Rx, at the Rx itself there is a long delay before detection. This is because the bacteria activate the pathways that generate green fluorescent proteins (GFP), which can then be used to indicate the arrival of MMs. In such  MC systems with long delays, it is demonstrated that by encoding information on the time between pulses, information rate can be significantly increased compared to CSK, especially by using techniques such as differential coding.

In \cite{Farsad2017_Communication} it was demonstrated that when information is encoded between the time of release of single MMs, and there is no ISI, the channel can be represented as an additive noise channel where the noise is distributed according to a stable distribution. For diffusion based propagation, this stable distributed noise has an infinite mean and variance, because of heavy tails. Hence geometric signal-to-noise ratio (GSNR) is introduced for stable distributed noises. The GSNR uses the geometric power of the noise $S_0(\mathscr{N})$, which is defined as 
\begin{align}
  S_0(\mathscr{N}) \triangleq e^ {E[{\log |\mathscr{N}|}]}  
\end{align}
where $|.|$ stands for the norm operator and $\mathscr{N}$ is the noise signal. It is shown that GSNR exhibits the same benefit as SNR in AWGN channel where the probability of bit error is the same for a given GSNR value regardless of the transmit or noise power \cite{Farsad2017_Communication, Gonzales2006_Zero}.

The final time-based technique changes the rate of release, i.e., the frequency of the release. This technique we call the Molecular Frequency Shift Keying (MFSK), which is similar to frequency shift keying (FSK), was first proposed and evaluated under idealistic settings in \cite{Mahfuz2010_Spatiotemporal}. A subsequent work \cite{Chou2012_MolecularCF} proposed a technique for detection of frequency modulated molecular signals. In particular, reactions-diffusion and reaction kinetic equations are used to model and design the reception process at the receiver. In the design of the receiver, ISI is also considered, and silence periods are used to mitigate ISI. The performance of the MFSK modulation is compared to CSK in \cite{Yilmaz2014_SimulationSO} using a particle based simulator and scenarios under which MFSK can achieve good performance are discussed. Finally, in \cite{Guo2015_MolecularLCW}, passband modulation is implemented by precisely controlling the longitudinal wave properties of MMs. 
Longitudinal waves are generated by the Tx oscillating towards or away from the Rx. This creates distinguishable carrier waves through a simple oscillation of the Tx, and can result in molecular frequency division multiplexing, where multiple transmitters can communicate with a single Rx with limited interference. 

\subsection{Spatial Techniques}

\begin{figure}[t]
  \centering
    \includegraphics[width=1.0\columnwidth]{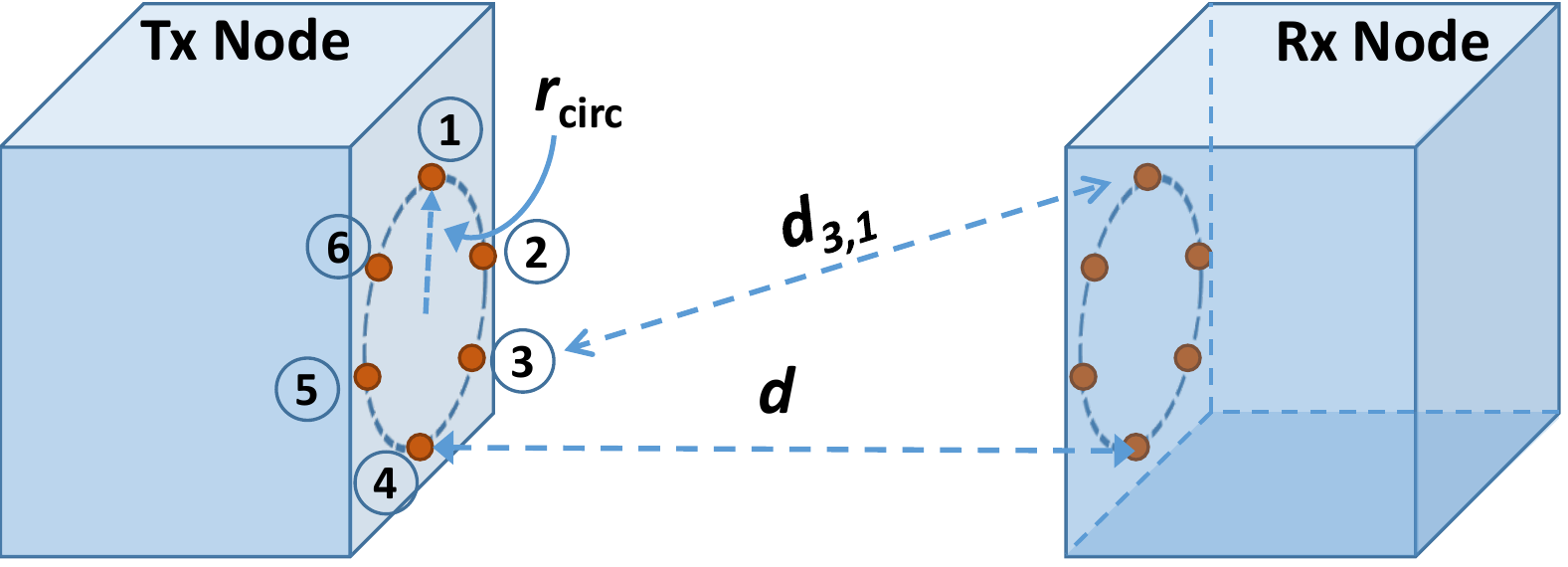}
  \caption{Example $6\times 6$ molecular MIMO system with uniform circular antenna (i.e., emission and absorption apparatus) placement. Indices of the antennas are shown in small circles starting from 1 to 6. In index modulation (IM), indices of the antennas represent the symbols and the Rx node detects the emission antenna index to demodulate the intended symbol. }
  \label{fig:modulation_index}
\end{figure}

Another group of techniques, called Spatial Techniques, aims to convey information by utilizing spatial diversity. Multi-antenna systems can embed the information into the emission location/antenna index and the Rx demodulates the symbol by estimating the emission location/antenna index. A $6\times 6$ example system is depicted in Fig.~\ref{fig:modulation_index} where the Tx and the Rx are $d$ apart and their antennas are placed in a circular way with a radius of $r_{\text{circ}}$. Please note that the distance between any given antenna pair (i.e., $d_{i,j}$: distance between the $\xth{i}$ antenna of the Tx and $\xth{j}$ antenna of the Rx) varies.

Spatial diversity is utilized in a simple way by considering molecular MIMO scenarios~\cite{meng2012mimoCB,koo2015detectionAF,lee2015molecularMIMO,koo2016molecularMIMO,damrath2017spatialDI,damrath2018arrayGA}. In these works, the achievable throughput is improved by utilizing multiple antennas with classical modulation techniques. Index modulation (IM), which utilizes indices of the transmitting antennas to convey additional information bits, was subsequently adapted to the MCvD system by Gursoy \etal{} who coined the term \emph{molecular IM}. In molecular IM, the message is modulated onto the index of emission antenna(s) and detected/demodulated by estimating the emission antenna index or indices. The most fundamental version of molecular IM, which only uses a single antenna, has been proposed as a new modulation technique called molecular space shift keying (MSSK)~\cite{gursoy2018indexMF}. Authors compared two parallel MSSK with Molecular Spatial Modulation (MSM) that utilizes MoSK for the symbol constellation (please note that both techniques use two types of molecules). The former technique is better for coping with ISI, on the other hand, MSM copes with Inter Link Interference (ILI) better.  In addition to MSSK, Gursoy \etal{} incorporated the Gray coding scheme for the antenna indices to reduce the BER. In~\cite{gursoy2019pulsePB}, Gursoy \etal{} combined MSSK with PPM technique and showed that this proposed combination technique achieved a significant improvement over both baseline MSSK and MSSK with CSK techniques. Spatial modulation is also analyzed in~\cite{huang2018spatialMF,huang2019spatialMF_TNBS} by considering selection combining and additionally equal gain combining in \cite{huang2018spatialMF} and maximum ratio combining in \cite{huang2019spatialMF_TNBS} at the Rx. In both studies, it is shown that the communication performance (in terms of BER) is improved significantly even with simple versions of index modulation, which only activates a single antenna for modulation at each symbol slot.

For $m$-ary MSSK, only one of the $m$ antennas ($a_0, a_1, ... a_{m\!-\!1}$) is active at each symbol slot, which reduces the ILI effect significantly. At the Tx, the antenna that emits the pre-determined amount of MMs, $A^{\text{Tx}}[k]$, is selected according to the value of the symbol as
\begin{align}
    A^{\text{Tx}}[k] = a_i, & \; \; \mbox{ where } S[k] = sym_{i}.
\end{align}
Then, the received symbol is demodulated as
\begin{align}
    \hat{S}[k] =  sym_\ell \;\;\;{\ell = \margmax\limits_{j\in \{0,...,m\!-\!1\}} N^{Rx,j}[k]},
\end{align}
where $N^{Rx,j}[k]$ denotes the total number of received molecules by the $\xth{j}$  antenna of the Rx (i.e., $a_j$) in the $\xth{k}$ symbol slot. In other words, the Rx performs maximum count detection. The possible error sources can be caused by ISI, ILI, and antenna misalignment. By checking the channel coefficients, it can be seen that the most prominent ILI-caused errors are due to the two adjacent Rx antennas of the intended one. Therefore, authors incorporated Gray coding for the antenna indexing to reduce the number of bit errors in connection with ILI~\cite{gursoy2018indexMF}. Even if the most basic version of molecular IM (i.e., MSSK) has promising results, the more generic molecular IM versions that allow multiple active antennas for MC have not been studied in the literature.

\subsection{Hybrid Techniques} 

As discussed in the previous subsections, most of the modulation techniques in the molecular communication literature focus on a single property of the molecular signal to vary according to the data to be transmitted. However, a handful of works propose techniques that utilize more than one of these properties. Although these techniques yield higher achievable bit rates, they require more complex mechanisms both at the Tx and at the Rx, which may limit their practicality.

In \cite{Kuran2010_Energy, Kabir2015_DMOSK}, two similar methods have been proposed that use multiple molecular signals, where these signals utilize CSK  but with different MM types ($mm_{type}$). From a conceptual point of view, this method, referred to as m-Channel CSK (m-CCSK) in this paper, where $m$ denotes the number of $mm_{type}$s being used, is similar to orthogonal channels in wireless communication. Considering a complete independency among the different $mm_{type}$s being used, the m-CCSK increases the achievable information rate by $m$, where each $mm_{type}$ is considered a parallel independent communication channel. However, such a technique requires the synthesis, storage, and release of $m$ MMs at the Tx, which will increase the energy cost of the communication by the same amount. At the Rx, there must be receptors suitable for each $mm_{type}$ covering the whole surface of the Rx which will make its design more complex and even impossible as $m$ increases.

Another technique that utilizes both the concentration and the type of MMs is the hybrid modulation scheme (HMS) that is proposed in \cite{Pudasini2014_RunLength}. Similar to the CSK-PA technique, HMS also leverages the fact that in some parts of the input signal there will be consecutive symbols having the same symbol value. As such, the number of consecutive symbols having the same symbol value becomes one input signal (i.e., repetition length signal - $\hat{l}[k]$) and the symbol value of each repetition becomes another input signal (i.e., symbol value signal - $\hat{V}[k]$). To give an example, considering a binary system for the following information
\begin{equation} \label{eq:HMS_Information}
S=\{1,1,0,0,1,0,1,1,1\},
\end{equation}
the two signals will be
\begin{align*} 
\hat{V}[k]&= \{1,0,1,0,1\}, \\
\hat{l}[k]&= \{2,2,1,1,3\}.
\end{align*}

By superimposing these two signals on two orthogonal molecular signal properties, where $N^{Tx}[k]$ is used for the repetition length signal, and $mm_{type}$ for the symbol value signal, the Tx is able to send a complex signal that carries many bits of information. As an example, in an HMS technique that uses two types of symbols and considers at most four consecutive symbols with the same symbol value (i.e., $HMS_{4,2}$), the reception ($\hat{V}[k]$, $\hat{l}[k]$) is defined as
\begin{equation} \label{eq:HMS_Rx_Decode_Symbol}
\hat{V}[k]= 
  \begin{cases}
    sym_0,       & N^{Rx}_{mm_{a}}[k] >  N^{Rx}_{mm_{b}}[k] \\
    sym_1,       & N^{Rx}_{mm_{b}}[k] \geq N^{Rx}_{mm_{a}}[k]
  \end{cases},
\end{equation}
and
\begin{equation} \label{eq:HMS_Rx_Decode_Length}
\hat{l}[k]= 
  \begin{cases}
    1, & \qquad\qquad N^{Rx}_{mm_{max}}[k] < \lambda_0 \\
    2,       & \quad\;\;\;\lambda_{0} \leq N^{Rx}_{mm_{max}}[k] < \lambda_{1}\\
    3,       & \quad\;\;\;\lambda_{1} \leq N^{Rx}_{mm_{max}}[k] < \lambda_{2}\\
    4,   & \quad\;\;\;\lambda_{2} \leq N^{Rx}_{mm_{max}}[k] \\
  \end{cases},
\end{equation}
where
\begin{equation} \label{eq:mm_max}
mm_{max}= 
  \begin{cases}
    mm_a,       & \hat{S}[k] = sym_0 \\
    mm_b,       & \hat{S}[k] = sym_1
  \end{cases}.
\end{equation}

As with m-CCSK, $HMS_{l,2^m}$ offers a much higher achievable data rate over CSK or MoSK at a cost of Tx and Rx complexity. The authors show that the highest achievable data rate gain can be attained where $m=1$ (i.e., each symbol represents 1-bit of information) and $l\in[2,20]$. However, such high $l$ values will necessitate a high number of thresholds, which will lead to high BER levels. The authors also proposed an alternative technique, HMS-D2, where $N^{Tx}[k]$ represents the symbol value signal and $mm_{type}$ represents the repetition length signal. This technique reduces the amount of threshold that is needed at the Rx, however, in this case the Tx will have to use $2-20$ different types of MMs which will increase the energy requirements at the Tx to impractical levels.

A different approach is to combine the type-based techniques with timing-based techniques. In \cite{Tang2020_Molecular}, authors propose molecular type permutation shift keying (MTPSK) which is in essence a combination of the CSK-SubTS and MoSK with ML detection. Here the system considers $M$ different type of molecules. Then, similar to the CSK-SubTS technique in a more generalized fashion, the $t_s$ is divided into $M$ sub-timeslots. At the first sub-TS, the Tx can choose to send any one of the $M$ molecule types. Then, in the second sub-TSs there will be $M-1$ options for selecting which molecule types to use. Therefore, for a given $M$ value, there will be $M!$ factorial permutations/symbols, and a symbol represents $\log_{2}{M!}$ bit(s) of information. As for the reception part, they have utilized an ideal ML detector as well as a practical implementation using a Viterbi-like algorithm. Using high signal power as well as selecting $M=8$, the proposed technique outperforms CSK, MoSK, as well as PPM. This improvement comes at a cost of high Tx and Rx complexity compared to these simpler techniques.

\subsection{Discussion on Appropriate Applications}
We have presented the comparison tables for each group of modulation techniques among the group itself, e.g., the concentration-based techniques are listed in Table~\ref{table:CSK-variants} without considering the cross comparison. Now, we will discuss the cross comparison in a qualitative and hypothetical framework with considering the challenges of MCvD system such as high ISI, synchronization, physical limitations, etc.

Concentration-based modulation techniques are, in general, simpler and may be more appropriate for less-capable devices/cells. When ISI mitigation capabilities are integrated to concentration-based modulations, the complexity of the technique increases directly. Therefore, less-capable devices should use simple concentration-based modulation techniques with long symbol duration to cope with ISI and they are appropriate for low-data rate applications or applications where a predefined message (e.g., alarm or sensed phenomenon) is sent only once.

\begin{figure*}[!t]
	\begin{center}
    \subfigure[Short $t_s$ ($2\,t_{\text{peak}}$) ]
		{\includegraphics[width=0.66\columnwidth,keepaspectratio]%
		{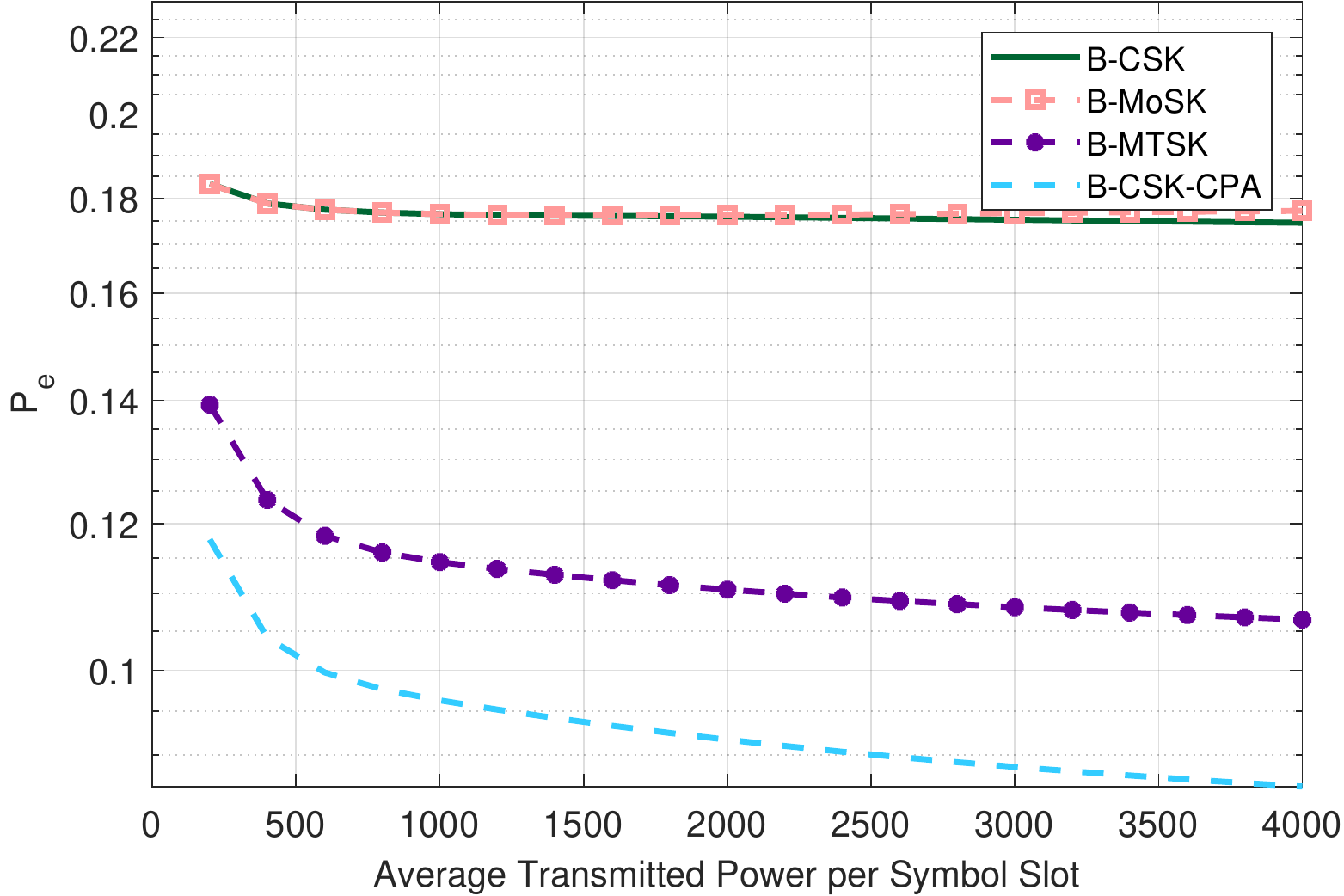}
		\label{fig_pe_cclass_short_ts}}
    \subfigure[Moderate $t_s$ ($2.5\,t_{\text{peak}}$) ]
        {\includegraphics[width=0.66\columnwidth,keepaspectratio]%
		{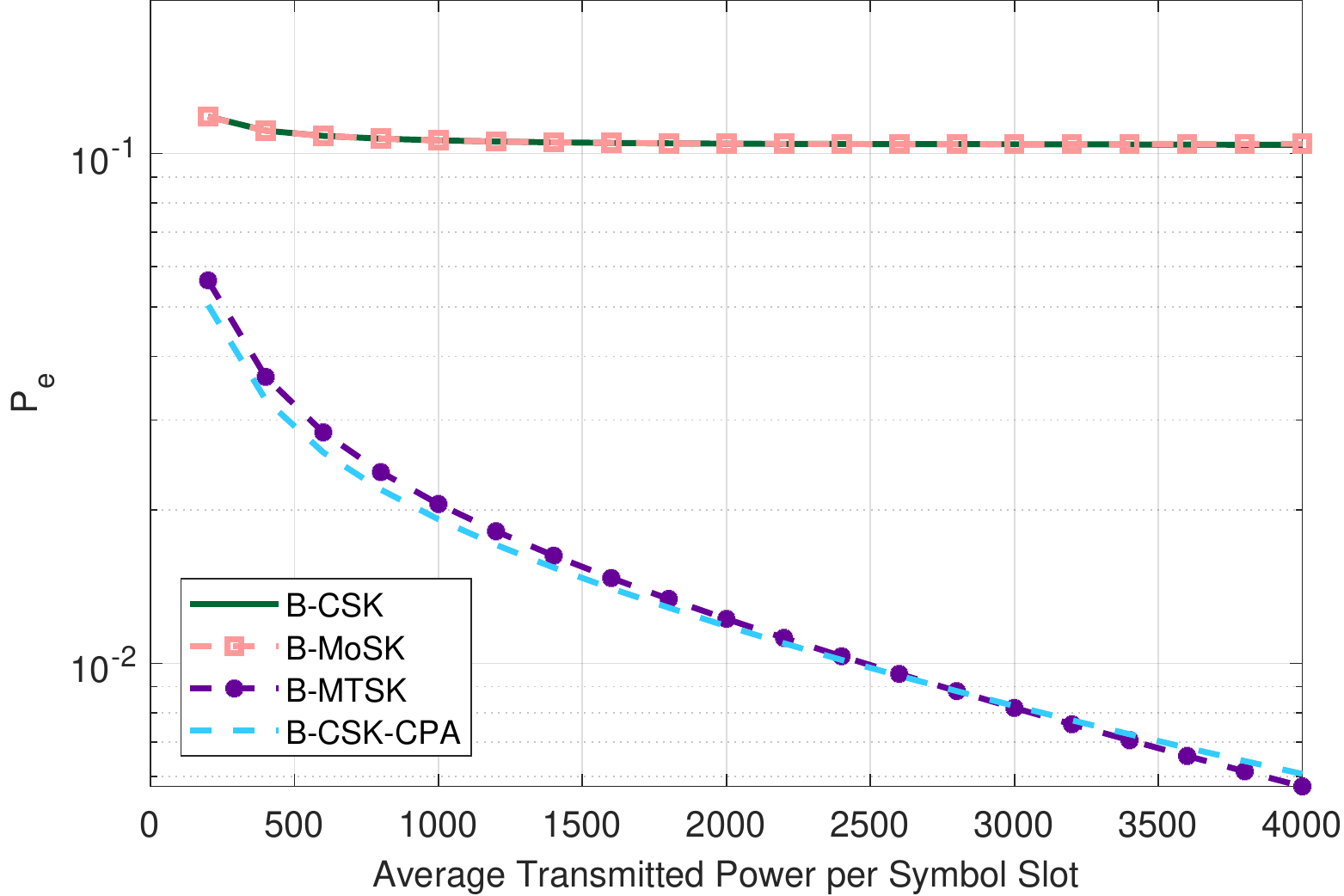}
        \label{fig_pe_cclass_moderate_ts}}
    \subfigure[Long $t_s$ ($3\,t_{\text{peak}}$) ]
        {\includegraphics[width=0.66\columnwidth,keepaspectratio]%
		{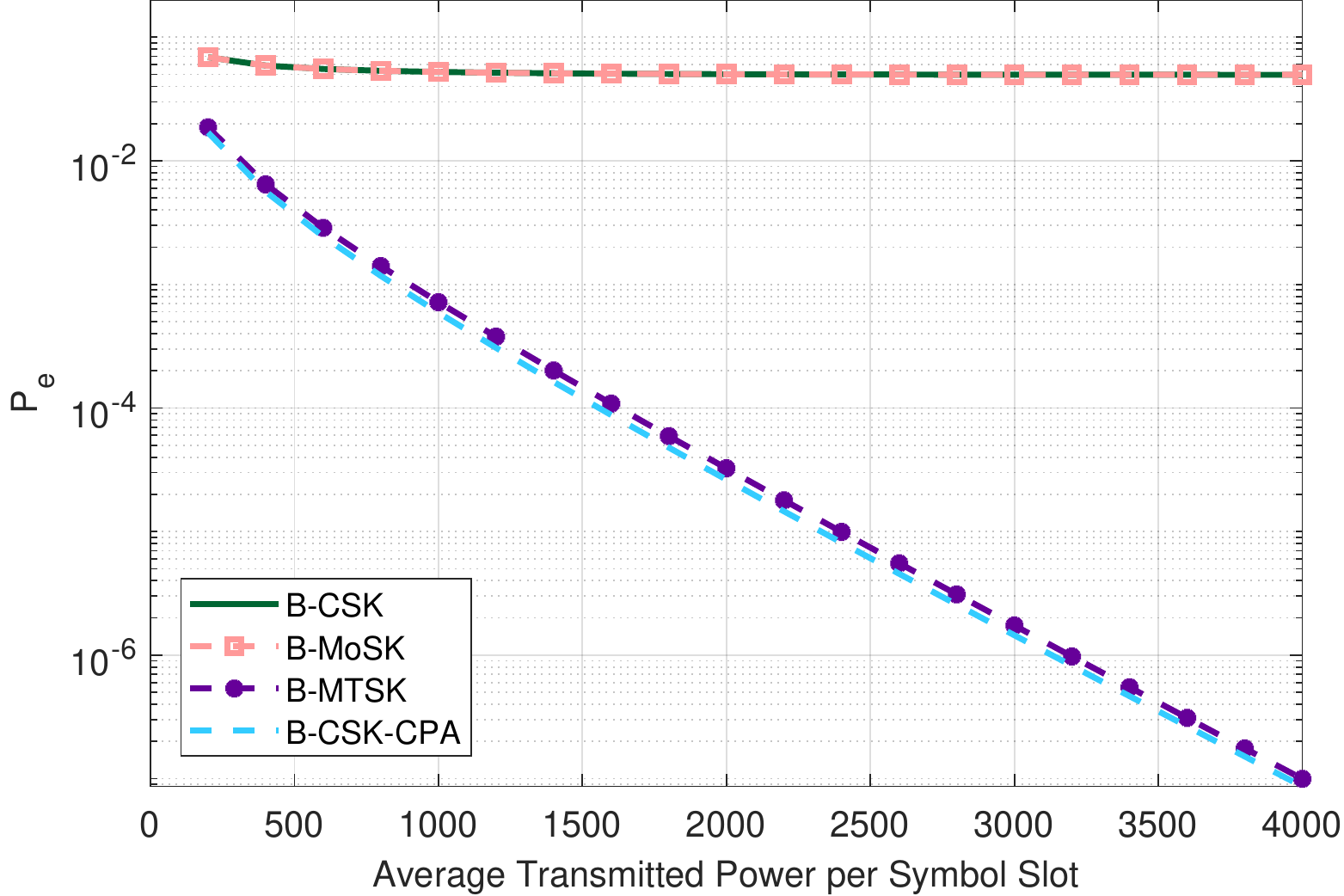}
        \label{fig_pe_cclass_long_ts}}
	\end{center}
   \caption{BER curves of Slow and Medium modulation techniques (i.e., B-CSK, B-MoSK, B-MTSK, and B-CSK-CPA). The $t_s$ values have been selected according to these categories.}
   \label{fig_pe_cclass}
\end{figure*}

Type-based techniques require different types of molecules for different symbols. Complexity of these techniques is higher than the concentration-based techniques. Hence, type-based techniques may not be the best choice for less-capable devices. However, if the device is capable of complex operations such as differentiating molecule types at the Rx side or synthesizing and sending different molecule types at the Tx side, then achieving ISI mitigation becomes easier compared to concentration-based techniques. Coping with ISI enables higher data rates. Therefore, these techniques may be promising for applications that are requiring higher data rates. Another advantage of using type-based techniques can be differentiating the nodes by assigning different molecule types to different links. However, this technique does not scale and may not be appropriate for crowded multi node topologies.

If we focus on the Timing-based techniques, they require accurate synchronization in time domain to decode the symbols accurately. These kind of modulation techniques are not appropriate for the application scenarios where the synchronization is hurdled. If the synchronization issue is handled adequately, then the channel capacity of the timing-based modulation technique scales at least polylogarithmically~\cite{Farsad2019_CapacityLimits}. Therefore, timing-based modulations have a great potential with a cost of synchronization requirements.

Spatial techniques are also promising for MCvD systems, however compared to the previous types they are less mature techniques and require further refinement. These techniques have potential to push the current data rate limits to even higher values. Due to having multiple antennas, such MCvD systems with spatial techniques can also be very appropriate for localization applications. However, they require advanced Tx and Rx structures with multiple emission and reception sites. Moreover, they may not be appropriate for crowded topologies due to multi-user interference and the required sensitivity for estimating the transmitting/emitter antenna.

\subsection{Performance Comparison}

\begin{figure*}[!t]
	\begin{center}
    \subfigure[Short $t_s$ ($1.75\,t_{\text{peak}}$) ]
		{\includegraphics[width=0.66\columnwidth,keepaspectratio]%
		{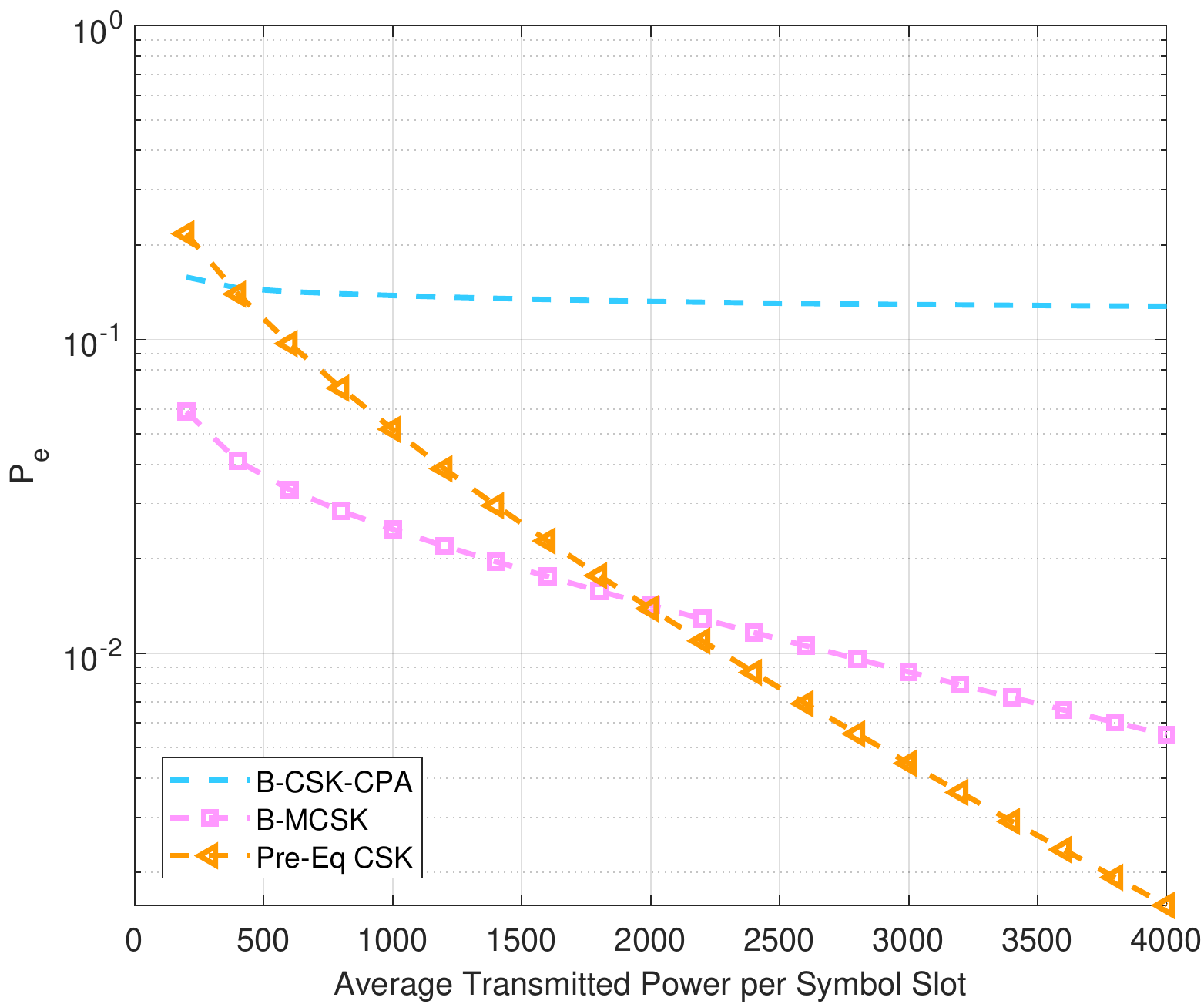}
		\label{fig_pe_gclass_short_ts}}
    \subfigure[Moderate $t_s$ ($2\,t_{\text{peak}}$) ]
        {\includegraphics[width=0.66\columnwidth,keepaspectratio]%
		{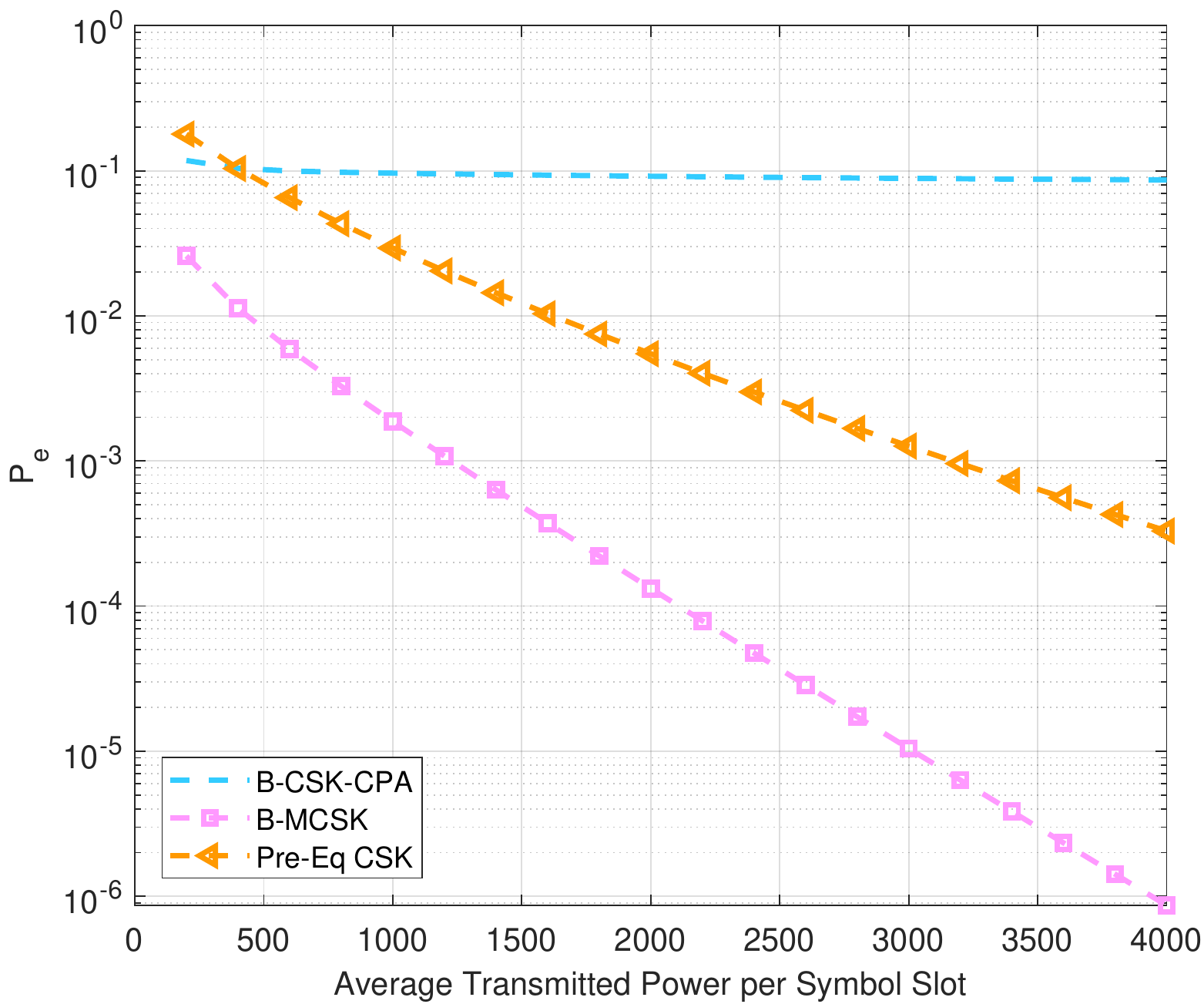}
        \label{fig_pe_gclass_moderate_ts}}
    \subfigure[Long $t_s$ ($2.25\,t_{\text{peak}}$) ]
        {\includegraphics[width=0.66\columnwidth,keepaspectratio]%
		{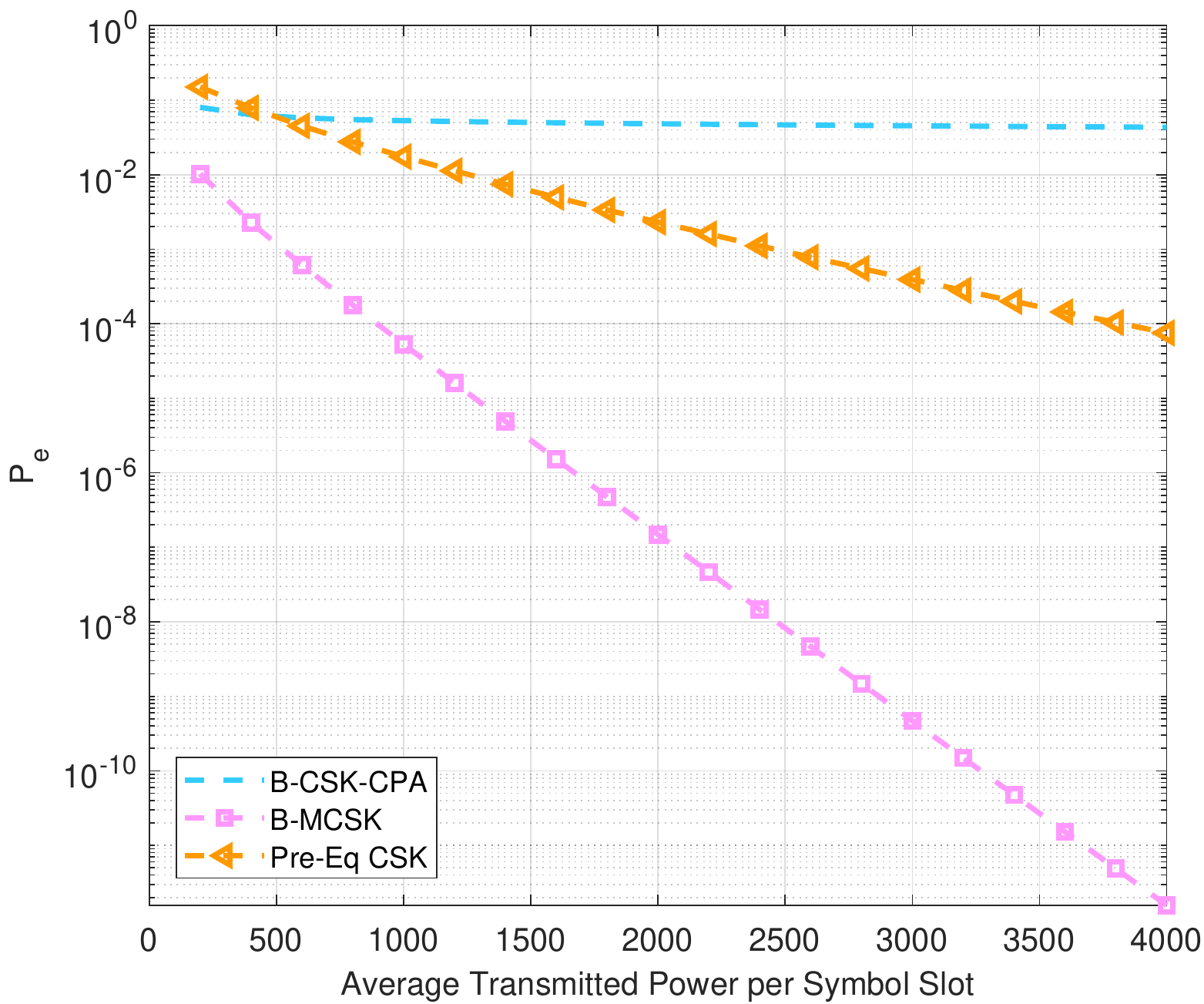}
        \label{fig_pe_gclass_long_ts}}
	\end{center}
   \caption{BER curves of Medium and Fast modulation techniques (i.e., B-CSK-CPA, B-MCSK, B-Pre-Eq. CSK). The $t_s$ values have been selected according to these  categories.}
   \label{fig_pe_gclass}
\end{figure*}

As we have now described the working principles of the modulation techniques in the literature, in this section we analytically evaluate the performances of key modulation techniques (i.e., CSK, MoSK, MTSK, CSK-CPA, MCSK, and Pre-Eq. CSK) in terms of BER for different average transmitted power values.  Regarding the MoSK technique, we utilize the scheme with the majority based detection variant. In order to have a fair comparison among these techniques, we normalized the $N^{Tx}[k]$ values for each symbol value so that the average transmitted power per symbol is the same for all the evaluated techniques. For example to compare the performances of B-CSK and B-MoSK techniques, where the average transmitted power is $500$, we set the $N^{Tx}[k]$ values as

\begin{equation} \label{eq:avg_power_example_CSK}
 N^{Tx}[k]_{\text{B-CSK}}=\quad
  \begin{cases}
    sym_0, & 0 \\
    sym_1, & 1000\\
  \end{cases},
\end{equation}
\begin{equation} \label{eq:avg_power_example_MoSK}
N^{Tx}[k]_{\text{B-MoSK}}= 
  \begin{cases}
    sym_0, & 500 \\
    sym_1, & 500\\
  \end{cases}.
\end{equation}

We use a 3D environment composed of a single point Tx device and a single fully absorbing spherical Rx device and without a drift component in the environment. Both the Tx and the Rx devices have a radius of $\SI{5}{\micro\metre}$ and the shortest distance between these two devices is also $\SI{5}{\micro\metre}$. The diffusion coefficient, $D$, has been selected as $\SI{69.4}{\micro\metre^2/\second}$ considering the viscosity of blood plasma at $\SI{37}{\celsius}$ \cite{freitas1999}. As for the MM, we use insulin molecules, whose Stokes radius is  $\SI{2.68}{\nano\metre}$ \cite{freitas1999}. 

The transmitted MC signal has a pulse waveform and we use the CIR calculation formula for 3D environments with a point source and fully absorbing receiver given in \cite{Yilmaz2014_3DChannel}. The BER performance has been evaluated where the effect of past $7$-symbols is taken into consideration (i.e., the ISI window length is $7$). We use a binary scenario, where each symbol represents a 1-bit information. As for the $t_s$ value, we use three different values as short $t_s$, moderate $t_s$, and long $t_s$. Each one of these values is given as multiples of the $t_{\text{peak}}$ value, which denotes the time of peak point of CIR function (i.e., $h(t)$) and it represents the time when the maximum number of MMs is received by the Rx device. This $t_{\text{peak}}$ value is also calculated by the aforementioned CIR formulation given the environmental parameters' values. 

The BER results of the chosen modulation techniques are given in Fig.~\ref{fig_pe_cclass} and Fig.~\ref{fig_pe_gclass}. In order to facilitate clear readability of the results, we categorize the modulation techniques in three categories as slow (i.e., CSK and MoSK), medium (i.e., MTSK, CSK-CPA) and fast (i.e., MCSK and PreEq CSK) techniques based on the $t_s$ values under which they exhibit a better BER performance. Fig.~\ref{fig_pe_cclass} depicts the result of the slow (i.e., CSK and MoSK) and medium (i.e., MTSK, CSK-CPA) techniques, while Fig.~\ref{fig_pe_gclass} shows the results of the medium and fast (i.e., MCSK and PreEq CSK) techniques.

As seen in all three subfigures of Fig.~\ref{fig_pe_cclass}, B-CSK and MoSK have similar performances after normalizing their average transmitted power values  regardless of the $t_s$ value. Also, in all three cases, both MTSK and CSK-CPA outperforms these fundamental techniques. As seen in Fig.~\ref{fig_pe_cclass_short_ts}, CSK-CPA outperforms MTSK in the short $t_s$ scenario. However, as seen in Fig.~\ref{fig_pe_cclass_moderate_ts} and Fig.~\ref{fig_pe_cclass_long_ts}, when the $t_s$ value increases, CSK-CPA loses this performance edge and its performance becomes similar to MTSK.

For the second group of techniques, due to their fast nature we use shorter $t_s$ values in all three cases. As seen in Fig.~\ref{fig_pe_gclass}, both MCSK and Pre-Eq CSK outperforms the best of the slow techniques in all $t_s$ values. Among the two fast techniques, Pre-Eq CSK gives a lower BER value considering a short $t_s$ value and where the average transmission power is greater than $2000$ (Fig~\ref{fig_pe_gclass_short_ts}). As $t_s$ increases though, MCSK outperforms the Pre-Eq CSK by a great margin both in the moderate $t_s$ and long $t_s$ scenarios. This is due to the fact that Pre-Eq CSK is in particular designed to operate under small $t_s$ values and the advantage of the equalization goes away as the $t_s$ value increases. One important take away from these results is the fact that MCSK not only has a much lower BER value than the other techniques considered, but it also has a moderate computational complexity. One potential drawback of MCSK is the synchronization requirement of the Rx, which might be a hard limitation in implementation.

Considering these results one can observe that the best modulation technique, in terms of BER, depends heavily on the design requirements of the target MC system, in particular the symbol duration, $t_s$, and the desired complexity. If a simpler system is desired clearly the CSK-CPA is the best choice since it is a concentration-based technique working with a single MM type. But it would necessitate a longer $t_s$ value to achieve a low BER. On the other hand, if it is possible to utilize multiple MMs, the question becomes what is the acceptable $t_s$ value which will have a serious impact over the system throughput. In case a short $t_s$ is desired, Pre-Eq CSK yields the lowest BER while if a longer $t_s$ will not cause too many problems MCSK clearly outperforms all other techniques. Though, as stated before MCSK requires stricter synchronization which may further complicate its adaptability in a real-life system.


\section{\label{sec:Open_Issues}Open Issues \& Future Directions}
This article has reviewed the modulation  techniques that have been developed for MC over the last decade. Since information in MC systems is carried by molecules, many new challenges in modulation design arise, which have been addressed in the recent literature. However, the field of MC is still in its infancy, with many open challenges remaining. In this section we briefly discuss some of these challenges along with relevant future research directions to address them.

Due to its nature, the received molecular signal inside a diffusion channel is inherently a positive valued signal. Therefore, methods that utilize both positive and negative components of a signal cannot be directly applied to the molecular communication. Similar to the molecular signal, the optical signal also exhibits such a behavior.  Consequently, modulation techniques inspired by optical systems are envisioned to be more suitable for MCvD systems. Some of the existing solutions can be applied to the problems that are encountered in MC domain but have similarity to already solved problems in optical systems.


\textbf{Synchronization} is one of the critical hurdles for modulation techniques in MC systems. As previously stated in Section~\ref{sec:Classification}, most of the works that propose modulation techniques for MC systems assume perfect synchronization between communication nodes, the Tx and the Rx. There are several studies proposing different synchronization techniques for MC systems by quorum sensing~\cite{abadal2011bioIS}, blind synchronization with channel delay~\cite{shahmohammadian2013blindSI}, and signal peak observation~\cite{mukherjee2018jointSA,jamali2017symbolSF,jamali2017symbolSF_TNB}. However, these methods are either complex for the nanomachines or they heavily rely on CSI - please note that acquiring CSI is a challenging task for time-varying MC channels. Moreover, the integration of these synchronization methods with the proposed modulation techniques is not investigated adequately. One important research direction is then the integration of synchronization methods with the existing modulation techniques and the investigation of asynchronous modulation techniques to increase the implementability of MC systems.

\textbf{Physical properties of the transmitter and the receiver} nodes in MC communications are critical for designing effective modulation techniques. Physical properties of the transmitter node determines the limitations on the emitted molecules. For example, if the transmitter node does not have enough storage, then the emission capabilities are limited by the storage capacity. Similarly, physical properties of the receiver nodes affect the CIR and CFRR, which also affects the modulation performance. In general, these physical properties and limitations are not considered adequately. For more realistic studies, these limitations should be considered in a holistic approach.

\textbf{Properties of the MC channel} is also critical for designing effective modulation techniques. MC channels possess slow propagation, high ISI, and channel memory, hence the channel significantly affects the modulation performance. Moreover, lack of a common SNR definition for the molecular communication systems is one of the critical issues for standardized comparison for modulation techniques. Much of the past MC research focused on ISI elimination and we discussed the modulation techniques that reduce ISI. Due to implementation challenges at small scales, new metrics can be proposed to compare modulation techniques while considering the complexity in a quantitative manner. For future directions, the tradeoff between ISI reduction and complexity can be analyzed in a detailed and quantitative manner. In addition, new modulation techniques can be introduced that are considering implementation issues (i.e., computational limitation of devices) more than the communication performance.

\textbf{Channel coding and precoding techniques} should also be considered to reduce the deteriorating effect of interference molecules. Precoding and channel coding techniques have potential to improve the reliability of the transmitted data. Moreover, long transmission duration suggest the use of error correcting codes instead of costly re-transmission strategies as in the case of satellite communications. The compatibility of the modulation and coding techniques is an open issue and should be investigated thoroughly.

All these MC specific challenges along with others such as multiuser techniques, networking, and potentially the physical layer security issues should be considered for future MC systems. In fact, the IEEE has established a standardization group (IEEE 1906.1) to address the various challenges of MC and develop standards around these solutions. Establishment of IEEE 1906.1 has the potential to push the technology towards commercialization if these challenges can be solved and adopted into a widely-used standard.

\section{\label{sec:Conclusion}Conclusion}
A key design component of the MC system is its modulation and associated demodulation mechanism. In the last decade or so a variety of modulation techniques have been proposed for the molecular communication in the literature. In this survey, we have summarized various modulation techniques and their tradeoffs. In order to analyze, categorize, and elaborate upon these techniques we provide a  framework that can also be utilized for future modulation techniques. Moreover, we provide a numerical performance comparison between the most prevalent of these modulation techniques and show that all of the techniques entail tradeoffs in different environments, with no single technique emerging as the best choice in all environments. Finally, we provide a brief elaboration on the open design issues about modulation techniques for MC and the future research directions to address them.


\ifCLASSOPTIONcaptionsoff
  \newpage
\fi



\bibliographystyle{IEEEtran}
\bibliography{Survey_Mod_Ench_2017}

%








\end{document}